\documentclass[11pt]{article}
\pdfoutput=1
\usepackage{amsmath}
\usepackage{amssymb}
\usepackage{graphicx,bbm,mathrsfs}
\usepackage{nicefrac}
\usepackage{slashed}
\usepackage{bbm}
\usepackage{geometry}
\usepackage{empheq}
\usepackage{stackrel}
\usepackage{ulem}

\usepackage{jheppub}
\usepackage{setspace}
\usepackage{relsize}

\usepackage{empheq}

\usepackage{mathtools}
\usepackage{enumitem}
\usepackage{dcolumn}   % needed for some tables
\usepackage{bm}        % for math
\usepackage{graphicx,mathrsfs}
\usepackage{nicefrac}
\usepackage{multirow}
\usepackage{color}
\usepackage{mathtools}
\usepackage{makecell}

\usepackage{environ} 
\usepackage{lipsum} 
 \NewEnviron{Smaller11}{
           \scalebox{1.1}{$\BODY$} 
 } 
 \NewEnviron{Smaller08}{
           \scalebox{0.8}{$\BODY$} 
 }

\newcommand{\be}{\begin{equation}}
\newcommand{\ee}{\end{equation}}
\newcommand{\bea}{\begin{eqnarray}}
\newcommand{\eea}{\end{eqnarray}}

\newcommand{\x}{\mathbf{x}}

\newcommand{\y}{{ r}}
\newcommand{\C}{c}
\newcommand{\B}{{\cal\bar B}}

\usepackage{wasysym}
\usepackage{hhline,colortbl}

\renewcommand{\S}{{\cal S}}

\newcommand{\nn}{\nonumber}

\renewcommand{\L}{L}

\newcommand{\isolated}{isolated }

\usepackage{titlesec}

\titleformat*{\section}{\Large\bfseries}
\titleformat*{\subsection}{\large\bfseries}
\titleformat*{\subsubsection}{\large\bfseries}
\titleformat*{\paragraph}{\large\bfseries}
\titleformat*{\subparagraph}{\large\bfseries}

\makeatletter
\newcommand*{\prodsym}{%
  \DOTSB
  \mathop{
    \mathchoice
      {\rlap{\kern.3em\rotatebox[origin=c]{-90}{}}{\prod}}
      {\vcenter{\rlap{\kern.2em\rotatebox[origin=c]{-90}{}}}{\prod}}
      {\sum}{\sum}
  }\slimits@
}
\makeatother

\makeatletter
\DeclareFontFamily{OMX}{MnSymbolE}{}
\DeclareSymbolFont{MnLargeSymbols}{OMX}{MnSymbolE}{m}{n}
\SetSymbolFont{MnLargeSymbols}{bold}{OMX}{MnSymbolE}{b}{n}
\DeclareFontShape{OMX}{MnSymbolE}{m}{n}{
    <-6>  MnSymbolE5
   <6-7>  MnSymbolE6
   <7-8>  MnSymbolE7
   <8-9>  MnSymbolE8
   <9-10> MnSymbolE9
  <10-12> MnSymbolE10
  <12->   MnSymbolE12
}{}
\DeclareFontShape{OMX}{MnSymbolE}{b}{n}{
    <-6>  MnSymbolE-Bold5
   <6-7>  MnSymbolE-Bold6
   <7-8>  MnSymbolE-Bold7
   <8-9>  MnSymbolE-Bold8
   <9-10> MnSymbolE-Bold9
  <10-12> MnSymbolE-Bold10
  <12->   MnSymbolE-Bold12
}{}

\let\llangle\@undefined
\let\rrangle\@undefined
\DeclareMathDelimiter{\llangle}{\mathopen}%
                     {MnLargeSymbols}{'164}{MnLargeSymbols}{'164}
\DeclareMathDelimiter{\rrangle}{\mathclose}%
                     {MnLargeSymbols}{'171}{MnLargeSymbols}{'171}
\makeatother

\begin{document}

\vspace*{4mm}

\thispagestyle{empty}

\begin{center}

%  {\LARGE
% \sc
\begin{minipage}{20cm}
\begin{center}
\hspace{-5cm }
\huge
\sc
Holography of  Linear Dilaton  Spacetimes \\
\hspace{-5cm } from the Bottom Up 
\end{center}
\end{minipage}
\\[30mm]

\renewcommand{\thefootnote}{\fnsymbol{footnote}}

{\large  
Sylvain~Fichet$^{\, a}$ \footnote{sylvain.fichet@gmail.com}\,, 
Eugenio~Meg\'{\i}as$^{\, b}$ \footnote{emegias@ugr.es}\,,
Mariano~Quir\'os$^{\, c}$ \footnote{quiros@ifae.es}\,
}\\[12mm]
\end{center} 
\noindent

${}^a\!$ 
\textit{Centro de Ciencias Naturais e Humanas, Universidade Federal do ABC,} \\
\indent \; \textit{Santo Andre, 09210-580 SP, Brazil}

${}^b\!$ 
\textit{Departamento de F\'{\i}sica At\'omica, Molecular y Nuclear and} \\
\indent \; \textit{Instituto Carlos I de F\'{\i}sica Te\'orica y Computacional,} \\
\indent \; \textit{Universidad de Granada, Avenida de Fuente Nueva s/n, 18071 Granada, Spain}

${}^c\!$  
\textit{Institut de F\'{\i}sica d'Altes Energies (IFAE) and} \\
\indent \; \textit{The Barcelona Institute of  Science and Technology (BIST),} \\
\indent \; \textit{Campus UAB, 08193 Bellaterra, Barcelona, Spain}

\addtocounter{footnote}{-1}

\vspace*{10mm}
 
\begin{center}
{  \bf  Abstract }
\end{center}
\begin{minipage}{15cm}
\setstretch{0.95}
\small
The linear dilaton background is the keystone of a string-derived holographic correspondence beyond AdS$_{d+1}$/CFT$_d$. This motivates an exploration of the $(d+1)$-dimensional linear dilaton spacetime (LD$_{d+1}$) and its holographic properties from the low-energy viewpoint. We first notice that the LD$_{d+1}$ space has simple conformal symmetries, that we use to shape an effective field theory (EFT) on the LD background. We then place a brane in the background to study holography at the level of quantum fields and gravity.  
    We  find that the holographic correlators from the  EFT feature  a pattern of singularities at certain kinematic thresholds. We argue that such singularities can be used to bootstrap the putative $d$-dimensional dual theory using techniques analogous to those of the Cosmological Bootstrap program. Turning on finite temperature, we study the holographic fluid emerging on the brane in the presence of a bulk black hole. We find that the holographic fluid is pressureless for any $d$ due to a cancellation between  Weyl curvature and dilaton stress tensor,    and verify consistency with the time evolution of the theory.   From the fluid thermodynamics, we find   a universal temperature and Hagedorn behavior for any $d$. This matches the properties of a CFT$_2$ with large $T\overline T$ deformation, and of little string theory for $d=6$. {We also find that the holographic fluid  entropy exactly matches the bulk black hole Bekenstein-Hawking entropy.}    Both the fluid equation of state and the  spectrum of quantum fluctuations suggest that the  $d$-dimensional dual theory arising from LD$_{d+1}$  is generically gapped.
    \vspace{0.5cm}
\end{minipage}

\newpage
\setcounter{tocdepth}{2}
\tableofcontents
\newpage

\section{Introduction \label{se:intro}}

The theory of quantum gravity likely encodes a holographic principle implying  that the information content inside any spacetime can be stored on its boundary~\cite{tHooft:1993dmi,Susskind:1994vu,tHooft:1999rgb,Bousso:1999xy,Bousso:2002ju}. 
The correspondence between AdS spacetimes and CFTs realizes very concretely this vision~\cite{Aharony:1999ti, Zaffaroni:2000vh, Kap:lecture,Nastase:2007kj}.  It further provides  the explicit description of the unitary evolution of boundary data.

Since the holographic principle itself applies to any spacetime, we might expect phenomena similar  to AdS/CFT to appear in other geometries, including flat and asymptotically flat spaces. 
The holographic dual theory of flat space remains elusive so far, although progress is being made via the study of celestial amplitudes, see {e.g.}~\cite{Pasterski:2021rjz} and references therein.  On the other hand, the holography of a specific asymptotically flat spacetime offers encouraging results. It is the so-called  $(d+1)$-dimensional \textit{linear dilaton} (LD$_{d+1}$) spacetime. 

The linear dilaton spacetime is a negatively curved  spacetime which is, in some sense, right in between AdS and Minkowski spacetimes. This fact can be recognized at the level of the entropy of massive black holes, which grows as  $E^{\frac{d-2}{d-1}}$,  $E^{\frac{d-2}{d-3}}$, and $E$ respectively for AdS, flat and LD spacetimes. In this work we  obtain the latter property for any $d$ via the properties of the LD holographic fluid. The fact that the LD space looks sometimes like flat space also appears in a variety of other ways throughout this work.

Below we briefly recall what is known about LD holography and then expose our approach.

\paragraph{LD holography from strings (review).}

It has long been known  that the linear dilaton background is  holographic  \cite{Aharony:1998ub}.  The string derivation of this fact involves a stack of NS5 branes taken in an appropriate decoupling limit.  It leads to a holographic correspondence between a string theory in a seven-dimensional (7D) LD  background and a 6D string theory on a stack of {NS5} branes with $g_s\to 0$ \cite{Aharony:1998ub}, where $g_s$ is the string coupling. The latter limit is known as little string theory (LST) \cite{Seiberg:1997zk,Berkooz:1997cq}. It is an interacting theory of noncritical strings which is nonlocal,  has no massless graviton, and has Hagedorn density of states at high energy \cite{Aharony:1999ks,Kutasov:2001uf}.   Lower-dimensional versions of this LD$_7$/LST$_6$ duality can be obtained via spatial compactifications. 

Even though we know that the dual LST {does exist},  it is  hard to study it because it seemingly has no independent formulation such as a Lagrangian description. 
However, it has been recently  proposed that a 2D compactification of LST with many fundamental strings may be described as a CFT$_2$ deformed by a specific single-trace $T\overline T$  deformation \cite{Giveon:2017nie}. As a result,  the 3D gravitational description of this deformed CFT$_2$ is  a string theory living on a background that interpolates between AdS$_3$ and LD$_3$. The LD$_3$ geometry corresponds to the large $T\overline T$ deformation regime. 
{The mass of some of these 3D spacetimes has been computed in \cite{Chang:2023kkq} and does seem to reproduce the effect of a $T \bar T$ deformation. }
Further evidence for the proposed duality involves the behavior of correlators and of certain symmetries (see \cite{Giribet:2017imm, Asrat:2017tzd,
Araujo:2018rho,Chakraborty:2020fpt,Georgescu:2022iyx,Chakraborty:2023mzc,Chakraborty:2023zdd,  Aharony:2023dod}, and \cite{Guica_lecture} for a review).

These progresses are obtained from the top-down via string models analyses.

\paragraph{LD holography from the bottom up. }

In this work we propose to explore the $(d+1)$-dimensional linear dilaton spacetime and its holographic properties from a bottom-up approach, working in the regime of subPlanckian energy scales. 
In this regime, gravity can be treated classically and effective field theory (EFT) techniques can be applied.\,\footnote{
Other field theoretical  studies  of  linear dilaton backgrounds
and related  phenomenological developments 
include Refs.~\cite{Antoniadis:2001sw, Antoniadis:2011qw, Csaki:2018kxb,Megias:2019vdb,Megias:2021mgj,Megias:2021arn,Antoniadis:2021ilm}. }

There is  an AdS/CFT analog to this approach. At the level of EFT, a version of AdS/CFT can essentially be derived by using symmetries and the conformal bootstrap 
\footnote{This leads to {a version of} the AdS$_{d+1}$/CFT$_d$ {correspondence} for any $d$ that we can summarize as: For any EFT in AdS there is a CFT, and for any CFT with large $N$ and with a large gap in the spectrum of higher-spin operators, there is an EFT in AdS.  }
(see {e.g.}~\cite{Heemskerk:2009pn,Fitzpatrick:2010zm, Kap:lecture,Caron-Huot:2021enk} as points of entry in the literature.)

In analogy with the EFT approach to AdS/CFT, our aim in the present paper is to explore some holographic aspects of the LD$_{d+1}$ background at low energies. The LD background has, of course,  less symmetry than the AdS one. Yet we will find that symmetries do play a role. 
The overall strategy is to place a brane in the LD$_{d+1}$ background, and to evaluate the correlators and the thermodynamic behavior as seen from this `braneworld'.

In this work we do {\it not} propose explicit dual theories for the LD$_{d+1}$ spacetimes.
While one may keep in mind that, for certain dimensions, the dual boundary theory might be some LST taken in an appropriate low-energy regime,  
our study remains agnostic of the current string knowledge about the boundary dual theories. 
 Our focus is on, \textit{i)} independently deriving familiar features from the low-energy viewpoint, and \textit{ii)}     uncovering new properties. We may wonder, for example, if some principle could be identified in order to bootstrap the $d$-dimensional dual theory.

\paragraph{Outline.}
In Sec.~\ref{se:LD_gen} we compute the linear dilaton background in any dimension, discuss its global properties, and identify symmetries of the LD line element. In Sec.~\ref{se:fields} we build an EFT for a scalar field  which is compatible with the  symmetries of the LD$_{d+1}$ background.
In Sec.~\ref{se:correlators} we place a {$d$-dimensional} brane in LD$_{d+1}$ and compute brane correlators on either side. 
In Sec.~\ref{se:finiteT} we allow a black hole solution in the bulk, 
compute the resulting holographic fluid on the brane and study its thermodynamic properties. Finally, Sec.~\ref{se:con} contains the detailed summary of our results, while App.~\ref{app:solutions} contains additional details on  the LD solutions with and without the bulk black hole.

\paragraph{Conventions.}

Throughout this work we use the conventions of Misner-Thorne-Wheeler  \cite{Misner:1973prb}, which include the mostly-plus  metric signature ${\rm sgn}(g_{MN})=(-,+,\ldots,+)$.
Likewise we  define the metric determinant $\sqrt{g}\equiv \sqrt{|\det{g_{MN}}|}$.  
The  number of spacetime dimensions is $D=d+1$, and we use either $D$ or $d$ depending on the context.

\section{Linear Dilaton Spacetime and  Symmetries
\label{se:LD_gen}
}

\subsection{Bulk Action}

We consider a $D$-dimensional spacetime whose coordinates are labeled as $x^M$.  
The general scalar-gravity action in the Einstein frame is
\begin{equation}
\S[g_{MN},\phi,\Phi] = \int d^Dx \sqrt{g} \left( \frac{M_D^{D-2}}{2} R - \frac{1}{2} (\partial_M \phi)^2 - V(\phi) +{\cal L}_m[\Phi,\phi]  \right)  \,,
% \nonumber \\
% - \int_{\textrm{brane}} d^{D-1}x \sqrt{\bar g}\, (V_{\brane}(\phi) + \Lambda_{\brane} ) + \ldots  \,.
\label{eq:action_bulk}
\end{equation}
where $R$ is the $D$-dimensional Ricci scalar, $\phi$ is the scalar field, $V(\phi)$ is the scalar potential, and  $M_D$ is  the fundamental $D$-dimensional Planck scale.   
The ${\cal L}_m$ Lagrangian contains matter fields, collectively denoted by $\Phi$, living on the spacetime background. The matter Lagrangian can  also depend on $\phi$. 

We assume that the scalar potential depends exponentially on $\phi$ as
\be
V(\phi) = -\frac{(D-2)^2}{2 }M_D^{D-2} k^2 \, e^{2\bar \phi}  \,\quad\quad {\rm with}\quad\quad \bar\phi = \frac{\phi}{\sqrt{D-2}\,M_D^{\frac{D}{2}-1}}\,, \label{eq:V}
\ee
where $k$ is a constant with mass dimension $1$ {and $\bar\phi$  is dimensionless}. 
This potential corresponds to a cosmological constant $-\frac{(D-2)^2}{2 }M_D^{D-2}k^2$ and no potential in the Jordan/string frame.  Throughout the paper  we refer to the scalar field $\phi$ as the \textit{dilaton}.

We split the $D$ coordinates as $x^M=\{x^\mu,\y\}$,  where $x^\mu = \{\tau,x^i\}$ are $D-1 \equiv d$ space-time coordinates, and assume a warped metric 
\be
ds^2=g_{MN}dx^Mdx^N= e^{-2A(\y)}\eta_{\mu\nu} dx^\mu dx^\nu+ e^{-2B(\y)}d\y^2\,.
\ee
Using this ansatz, the independent field equations  for the metric and the scalar are
\begin{eqnarray}
\hspace{-0.3cm}&& A^{\prime\prime}(r) + A^\prime(r) B^\prime(r) - \bar \phi^\prime(r)^2 = 0 \,,  \label{eq:EoM1}  \\
\hspace{-0.3cm}&&A^\prime(r)^2 + \frac{1}{D-1} \left(   2 e^{-2 B(r)} \bar 
V(\bar\phi) - \bar\phi^\prime(r)^2 \right) = 0 \,,    \label{eq:EoM2}
\end{eqnarray}
where $\bar V\equiv V/[(D-2) M_D^{D-2}]$ {has mass dimension 2}. There is a third field equation which is redundant, as it can be expressed in terms of these two equations (see Eq.~(\ref{eq:identity})).

\subsection{The Linear Dilaton Spacetime}

A canonical solution to the field equations \eqref{eq:EoM1}-\eqref{eq:EoM2} is
\begin{align}
ds^2_{\rm LD} &= d\y^2+\frac{\y^2}{\L^2}\eta_{\mu\nu}dx^\mu dx^\nu \,, \label{eq:ds2_LD} \\
\bar \phi(r) &= -\log \left( k\y\right)  \,
\end{align}
{where $L$ is an integration constant with dimension of length. The  field equations \eqref{eq:EoM1}-\eqref{eq:EoM2} have a total of three integration constants.  All the solutions are equivalent to Eq.\,\eqref{eq:ds2_LD}
up to coordinate transformations. 
The complete set of solutions, as functions of the integration constants {$a$, $\C$ and $L$}, is presented in App.\,\ref{app:solutions}. }

 The Einstein tensor obtained from the $ds^2_{\rm LD}$ metric  is
\be
G^{\rm LD}_{\mu\nu}=  \frac{(D-2)(D-3)}{2\L^2}\eta_{\mu\nu} 
\,,\quad \quad G^{\rm LD}_{\y\y}= \frac{(D-1)(D-2)}{2\y^2}  \,,\quad\quad G_{\mu \y}=0\,,
\ee
and the Ricci scalar is 
\be
R_{\rm LD}=  -\frac{(D-1)(D-2)}{\y^2}   \,.\label{eq:R_LD}
\ee
There is a curvature singularity located at $r \to 0  $ for $D>2$.

Throughout the paper, we sometimes use the conformal coordinates $(x^\mu,z)$, $z= L \log\frac{L}{\y}$. The scalar field varies linearly in $z$  in conformal coordinates, which is why the spacetime was dubbed   ``linear dilaton''.

\subsubsection{$D=2$ dimensions}

We briefly comment  on the case of $D=2$ spacetime.  We notice that  for $D=2$ the solution simply reduces to a flat space metric since $R_{MN}$ and $V(\phi)$ vanish identically. We can wonder to which region of flat space it maps exactly. To answer this, we go to Cartesian coordinates. 
For Lorentzian signature $ds_{\rm LD}^2=d\y^2-\frac{\y^2}{L^2} dt^2$, the mapping to coordinates $(x,\tau)$ is given by {$r=x$,} $t=L \tanh^{-1}(\frac{\tau}{x})$. 
This implies $x^2\geq\tau^2$, therefore we obtain that the metric maps to 2D flat spacetime restricted to \textit{spacelike} intervals.  
In other words, in $D=2$ the $ds_{\rm LD}^2$ metric amounts to a two-dimensional version of the Rindler metric. Our focus in this work is   on $D>2$.

\subsubsection{Global features}

\paragraph{Singularity.}
The curvature singularity appearing at $\y\to0$ for the LD
solutions can be classified as ``good'' in the sense of Refs.~\cite{Gubser:2000nd,Cabrer:2009we}. 
We will explicitly see the expected physical properties of the good singularity:  It repels quantum fluctuations  (Sec.~\ref{se:correlators}) and gets censored by a horizon (Sec.~\ref{se:finiteT}).

\paragraph{Boundaries.}  By going to conformal coordinates with $z= L \log\frac{L}{\y}$, we can see that the LD line element is proportional to  $ds^2\propto dz^2 +\eta_{\mu\nu}dx^\mu dx^\nu$ with $(x^\mu,z)\in {\mathbb{ R}}^{1,d}$. 
{Thus the conformal structure of the LD spacetime is a null diamond just like Minkowski spacetime (and unlike  AdS).}~\footnote{
{For example, the Poincaré patch of AdS$_{d+1}$  is conformal to ${\mathbb{ R}_+ \times \mathbb{ R}}^{1,{d-1}}$ hence its Penrose diagram is a half-null diamond. }} Notice that this is in spite of the fact that the scalar curvature and the potential blow up at $z=\infty$.

\paragraph{Geodesics.}

Following \cite{geodesic_note}, the timelike geodesics  are attracted towards the singularity, and the geodesic distance grows exponentially with the separation  $L \sqrt{\eta_{\mu\nu}(x_1^\mu-x_2^\mu)(x_1^\nu - x_2^\nu)}$.
In contrast, the null geodesics escape the singularity.  In conformal coordinates they behave as straight lines analogously to flat space.\,\footnote{We thank S. Barbosa (S. Fichet's master student) for sharing his notes.}

\subsection{Symmetries} \label{se:LD_symmetries}

What are the symmetries of the line element of the LD spacetime, Eq.\,\eqref{eq:ds2_LD}? The line element has  $d$-dimensional Poincaré isometries along the constant-$\y$ slices.   It has other symmetries though.
\begin{itemize}

 \item   The line element has a manifest {conformal symmetry}: the dilatation 
   \be \y\to \lambda \y \,,\quad x^\mu\to x^\mu \label{eq:D}\ee 
   which gives $ds_{\rm LD}^2\to \lambda^2 ds_{\rm LD}^2$. That is, under dilatation the line element is equivalent to itself up to a constant Weyl transformation.
   
 \item The line element has  a manifest discrete ($\mathbb{Z}_2$) {conformal  symmetry}: the inversion 
 \be \frac{\y}{\L}\to  \frac{\L}{\y} \,, \quad x^\mu\to x^\mu
 \label{eq:inversion}
 \ee 
which gives \be ds_{\rm LD}^2\to \frac{\L^4}{\y^4} ds_{\rm LD}^2\,.
 \label{eq:ds2LD_inversion}
 \ee 
 That is,  under inversion the line element is equivalent to itself up to a Weyl transformation. 
\end{itemize}

The discrete conformal symmetry of $ ds_{\rm LD}^2$ is expressed here as an  inversion in the  $(x^\mu,\y)$ coordinates, but of course, the existence of a symmetry is coordinate-independent.  For exam\-ple, in conformal $z=\L \log \frac{L}{\y}$ coordinates, this symmetry is a ``conformal parity'', {i.e.}~the line element is equivalent to itself under $z\to-z$ up to a Weyl transformation. 
The coordinate-independent statement is that the {{LD line element is conformally invariant under an involution map}}.   In the rest of the paper we mostly use $\y$ coordinates and will refer to the conformal $\mathbb{Z}_2$ symmetry of the line element as the ``conformal inversion symmetry''.

\subsubsection{Symmetry as  a defining property}
\label{se:sym_prop}

Let us  study to which extent the peculiar conformal symmetries identified above uniquely define the linear dilaton spacetime.  
Poincaré invariance imposes the warped ansatz $ds^2= F(\y)d\y^2+G(\y) \eta_{\mu\nu}dx^\mu dx^\nu$. 
Then requiring that  dilatation in $\y$ leaves the line element conformaly invariant imposes the form {$ds^2= F(\y)(d\y^2+(\y^2/L^2) \eta_{\mu\nu}dx^\mu dx^\nu)$}. Such a metric is more general than the LD one. Further requiring that the conformal weight be a constant, {i.e.}~that the line element be \textit{homothetic} to itself, we obtain the form
\be 
ds_{a}^2 = \frac{\y^{2a}}{L^{2a}}  \left(  \frac{L^2}{r^2} d\y^2 + 
 \eta_{\mu\nu}dx^\mu dx^\nu \right) \,.
\label{eq:ds2a}
\ee
For $a\in\mathbb{R}_{/\{0\}}$,  $ ds_{a}^2$  reproduces exactly the line element (\ref{eq:dsa2}), which is   equivalent to the LD line element 
 Eq.\,\eqref{eq:ds2_LD} via the coordinate transformation  $ \frac{\y}{L} \to \frac{1}{a} \left( \frac{\y}{L} \right)^{a}$, $x^\mu \to a x^\mu$ (see {App.~\ref{subsec:sol_T0}}). The remaining case $a = 0$ corresponds to Minkowski spacetime, in which case the scale transformation  amounts to translation  along the  dimension $x_{d+1}\equiv L \log\frac{\y}{L} $. In this particular case the dilatation in $\y$ becomes an isometry.

The line element $ds^2_a$ has the conformal inversion symmetry for any $a$. For $a\in\mathbb{R}_{/\{0\}}$,  $ ds_{a}^2$ is the one of the LD spacetime, while for $a=0$ it reduces to standard parity along the  dimension $x_{d+1}\equiv L \log\frac{\y}{L} $. One can easily show that it is the unique involution in $\y$ of the $ds^2_a$ metric. 

Finally, one may ask whether the existence of a conformal symmetry under an involution in $\y$   restricts by itself the warped ansatz to be LD: The answer is negative.\,\footnote{For example, consider the set of metrics {$ds^2 = d\y^2 + c^2(\y)\eta_{\mu\nu}dx^\mu dx^\nu$} with $c(\y)$ an arbitrary real function, and consider an involution map  $\rho$, defined by $\rho\circ\rho=$ Id, acting on the $\y$ coordinate. We find that choosing $\rho$ such that  $c(\y) \rho'(\y) =c(\rho(\y))$ gives a metric which is conformally equivalent to itself under the $\rho$ involution. {This is much more general than the LD spacetime.}  } 
Extra assumptions are needed to reach the LD spacetime by only imposing symmetry under an involution. This happens for example for a line element $ ds^2= \frac{\y^{2a}}{L^{2a}}\left( \frac{L^2}{\y^2} d\y^2+\frac{\y^{2(b-2)}}{L^{2(b-2)}} \eta_{\mu\nu} dx^\mu dx^\nu\right)$, for which conformal invariance under inversion requires $b=2$, which yields 
Eq.\,\eqref{eq:ds2a}.

In summary,  requiring that the warped ansatz be homothetic to itself under dilatation of $r$ gives rise to either  the linear dilaton spacetime or the Minkowski spacetime. The inversion symmetry automatically follows. In short, we can say that there is a symmetry which uniquely defines the linear dilaton spacetime. 
We have checked that imposing only the inversion symmetry  gives a weaker constraint.

\subsection{Effective Field Theory and  $S$-duality}

For $D\geq 4$ the theory must be considered as a low-energy effective field theory (EFT) describing gravity at subPlanckian scales. The Einstein-Hilbert action is then the leading order term in a curvature expansion of the quantum effective action. 
The EFT Lagrangian in the gravity sector reads in general
\be
\frac{1}{\sqrt{g}}{\cal L}_{\rm eff}= \frac{1}{2}M_D^{D-2} R + \sum^\infty_{n=2} \frac{a_{n,i}}{\Lambda^{2n-D}}{\cal O}_{n,i} \,, \label{eq:Leff}
\ee
where $\Lambda$ is the scale at which the effects of the UV-completion of quantum gravity show up. From the viewpoint of the low-energy gravity EFT, it is the typical cutoff scale below which the EFT is valid. 
   The ${\cal O}_{n,i}$ are local operators made of $n$ curvature tensors, {e.g.} ${\cal O}_{2,i}=\{R^2, (R_{MN})^2, (R_{MNPQ})^2\}$ and their derivatives.  
Such corrections happen if any state with mass of order $ \Lambda$ is integrated out in the UV completion.

\subsubsection{Validity domain of the linear dilaton spacetime}

The EFT breaks down when the series in Eq.\,\eqref{eq:Leff} cannot be truncated, {i.e.}~when the $R/\Lambda^2$ expansion breaks down.
This validity cutoff of the EFT is reached at {e.g.}~high curvatures or small distances. Since in the LD spacetime the curvature depends on the location in the bulk, the EFT cutoff  depends on the location $\y$ in the bulk. (Notice this is a general feature of EFT in curved space, which happens even for AdS \cite{Fichet:2019hkg,Costantino:2020vdu}).

In the LD metric Eq.\,\eqref{eq:ds2_LD}, and using the corresponding  curvature tensor Eq.\,\eqref{eq:R_LD},   we can see that the domain of validity of the EFT is 
\be
\y\gtrsim \frac{1}{\Lambda} \,, \label{eq:r_cond}
\ee
where we have assumed that the $a_{n,i}$ coefficients are $O(1)$. 

The bound \eqref{eq:r_cond} implies that the inversion symmetry is supported on the restricted domain $\y \in [\frac{1}{\Lambda},\Lambda L^2]$. This interval is non empty if the cutoff scale satisfies 
\be
\Lambda > \frac{1}{L}\,.\label{eq:Lambda_cond}
\ee
The inversion symmetry becomes immaterial if this condition is not satisfied.

\subsubsection{
On inversion symmetry and $S$-duality
}

We notice that the  inversion \eqref{eq:ds2LD_inversion},  
a conformal symmetry of the metric, transforms the dilaton expectation value as 
\be
\phi_{\rm LD}\to - \phi_{\rm LD} + {\rm cte}\,. 
\ee
Such a transformation might be  reminiscent of the  $S$-duality of string theory (see e.g.~\cite{Tong_lecture,Wray_lecture} for introductions).

The linear dilaton spacetime, at least in certain dimensions, can be thought of as a compactified limit of type IIB string theory, which is self-dual under $S$-duality \cite{Aharony:1999ti,Itzhaki:1998dd}. In this case the $S$-duality identifies two solutions with each other, hence defining a $\mathbb{Z}_2$ symmetry.
It is thus not surprising to find  in our framework two solutions which are identified to each other via an $S$ duality-like transformation. 

The string $S$-duality exchanges weak and strong coupling, $g_s \to 1/g_s $. In our EFT this notion maps  onto  the strength of  the higher order curvature corrections to gravity, which varies with $\y$.  The region of strong coupling is the one toward the singularity,
where the EFT of gravity breaks down.  The region of weak coupling is the one away from the singularity, where the EFT is valid. 
The $S$-duality  suggests that  there should be an alternative weakly coupled description beyond the  region where the EFT validity breaks down, $\y<\frac{1}{\Lambda}$. This is, in a sense, what we observe: there is indeed a weakly-coupled region at small $\y$, namely $\y<\Lambda L^2$, which corresponds to the original solution transformed under \eqref{eq:ds2LD_inversion}. 

In the following we do not need to refer to $S$-duality, and we denote  the inversion symmetry as $S$ when acting on the fields.

\section{Fields in a Linear Dilaton Background }

\label{se:fields}

In this section we consider quantum fields living on the linear dilaton background. These are contained in the matter Lagrangian in Eq.\,\eqref{eq:action_bulk}.
The quantum fields inherit the symmetries of the linear dilaton background. We study the effects of these symmetries on the fields and how they constrain physical observables. For concreteness we focus on a real scalar field $\Phi(x^M)$ and its correlators.

\subsection{Symmetries and Free Fields}

The fields living in the LD background Eq.\,\eqref{eq:ds2_LD} are constrained by Lorentz invariance along the constant-$r$ slices, thus $\Phi=\Phi(x_\mu x^\mu,\y)$. Moreover they should transform as representations, in field space, of the conformal dilatation, Eq.\,\eqref{eq:D}, and the conformal inversion,  Eq.\,\eqref{eq:inversion}, symmetries.

The fundamental action of the matter fields on the LD 
%and LD$^\prime$ 
background is denoted as 
\be
\S_{\rm LD^{}}[\Phi]\equiv
\S[g^{{\rm LD^{}}}_{MN},\phi^{{\rm LD^{}}},\Phi] \,.
\ee
%and similarly for the LD$^\prime$ background. 
%
Let us consider a free field with canonical normalization,
\be
\S^{\rm free}_{\rm LD^{}}[\Phi]=-\frac{1}{2}\int d^dx\, d\y \frac{\y^d}{L^d} \partial_M\Phi \partial^M\Phi  \,. \label{eq:Sfree}
\ee
The corresponding d'Alembertian is defined by
\be
\square_\y = \frac{L^d}{\y^d}{\partial}_\y \left(\frac{\y^{d}}{L^{d}} \partial_\y\right) + \frac{L^2}{\y^2}\square^{(d)} \,,
\ee
with $\square^{(d)} = \eta^{\mu\nu}\partial_\mu\partial_\nu$ the $d$-dimensional flat space d'Alembertian.

\subsubsection{Dilatation}
\label{se:dilatation_field}

Since the LD background has conformal dilatation invariance in $\y$, any operator $\cal O$ on the LD background should transform {under a dilatation transformation $D_\lambda$} as 
\be
D_\lambda [{\cal O}(x,\y)]={\cal O}(x,\lambda \y) = \lambda^{-\Delta_{\cal O}} {\cal O}(x,\y) \,,
\ee
where we refer to $\Delta_{\cal O}$ as the scaling dimension. 

How should a free field scale upon dilatation in $\y$ of the LD background? 
The condition  is that the action of the free field should be invariant under the dilatation, 
{
\be D_\lambda[{\S}_{\rm LD}]={\S}_{\rm LD}
\label{eq:DSLD}
\,.\ee 
}
This condition is needed to ensure that the correlators ({i.e.}~the derivatives of the partition function)  transform as representations of dilatation. This would not be the case if the action transformed nontrivially under dilatation.
 Using the transformation $D_\lambda [\Phi(\y)] = \lambda^{-\Delta_\Phi}\Phi(\y)$,
  the scaling dimension of the free field  must be 
  \be \Delta_\Phi = \frac{d-1}{2}\,\label{eq:Delta_Phi} \,.\ee
  
{The result \eqref{eq:Delta_Phi} is also independently verified via a 
 direct calculation of the 2-point (2-pt) function.}
As a side note, in our conventions the dimension Eq.\,\eqref{eq:Delta_Phi} happens to match  the standard mass dimension of fields in flat space.

\subsubsection{Inversion}

We have seen that the line element of the background is self-equivalent under the inversion operation \eqref{eq:inversion}  up to a conformal factor. This transformation can be equivalently thought of as a Weyl transformation {i.e.}~a redefinition of the metric field. We may thus expect that operators on the LD background transform under the inversion up to a similar redefinition. The transformed operator $S[{\cal O}]$ should transform under dilatations, thus the field redefinition should be a power of $\y$, 
\be S[ {\cal O}(x,\y)] =  {\cal O}\left(x,\frac{L^2}{\y}\right) \frac{\y^c}{L^c} \,,\ee 
{where $c\in \mathbb R$ is a constant.}
We have that \be
S^2 [{\cal O}] = S\left[ {\cal O}\left(x,\frac{L^2}{\y}\right) \frac{\y^c}{L^c}\right] = {\cal O} \,,
\ee
therefore $S$ is an inversion. 
Requiring that the matter action be invariant under $S$, we find that the free field transforms as
\be
S[\Phi(x^\mu, \y)]=\Phi\left(x^\mu, \frac{L^2}{\y}\right) \frac{L^{d-1}}{\y^{d-1}}\equiv \hat \Phi(x^\mu, \y)\,,
\ee
{i.e.}~$c=1-d$ for the free field. 

The following identities hold: \be \square_{\frac{L^2}{\y}}= \frac{\y^4}{L^4} \square_{\y} \Big|_{d\to 2-d}\,,\quad\quad 
\square_{\frac{L^2}{\y}} \Phi\left(\frac{L^2}{\y}\right) = 
\frac{\y^{d+3}}{L^{d+3}} 
\square_{\y} \hat\Phi(\y) \,.
\ee
The invariance of the action is then verified by using
\begin{align} 
{\S}^{\rm free}_{\rm LD}[ \Phi] & = \frac{1}{2} \int  d^dx\, d\y
\frac{\y^d}{L^d}\Phi(x^\mu,\y)\square_\y \Phi(x^\mu,\y)
\\
& = \frac{1}{2} \int  d^dx\, d\y
\frac{L^{d+2}}{\y^{d+2}}\Phi\left(x^\mu,\frac{L^2}{\y}\right)\square_{\frac{L^2}{\y}} \Phi\left(x^\mu,\frac{L^2}{\y}\right)
 ={\S}^{\rm free}_{\rm LD}[\hat \Phi] \,,
\label{eq:SSLD}
\end{align} 
where we used the change of variable {$r/L\to L/r$} 
%$\y\equiv\frac{L^2}{\y^\prime}$ 
in the second line.

\subsubsection{Mass}

\label{se:Mass}

A mass term is similarly constrained by the dilatation and inversion symmetries.  We find
\be
{\cal S}_{\rm LD}^{\rm mass}= -\frac{1}{2}\int d^dx\, d\y \frac{\y^{d-2}}{L^{d-2}} m^2 \Phi^2 \,,
\label{eq:Lm}
\ee
where $m$ is constant. Notice the nontrivial scaling in $\y$.

{We may also  write the mass term (\ref{eq:Lm}) in a covariant form that distinguishes the metric factor $\sqrt{g}=(r/L)^d$ from the dilaton background. In that view the mass can be understood as a function of the dilaton $\phi$, with $m(\phi)\equiv kL e^{\bar\phi} m $. 
This form is not needed for our EFT analysis of a matter field living on the LD background.}

We can also notice that if we use a non-canonical field {$\Phi\equiv \frac{\y^u}{L^u} \tilde \Phi$} for any $u\in\mathbb{R}$, the kinetic term 
Eq.\,\eqref{eq:Sfree} expressed in $\tilde\Phi$ generates a mass term with precisely the form of Eq.\,\eqref{eq:Lm}. This ensures that the structure of the Lagrangian is not spoiled if one uses non-canonical fields --- the symmetries always constrain the mass term to a unique monomial in $r$.

The  mass term Eq.\,\eqref{eq:Lm} together with the d'Alembertian form the wave operator
\be
{\cal D} = -\square_\y +\frac{L^2}{\y^2}m^2 = -\frac{L^d}{\y^d}{\partial}_\y \left(\frac{\y^{d}}{L^{d}} \partial_\y\right) + \frac{L^2}{\y^2}\left(-\square^{(d)}+m^2\right) \,.
\label{eq:D_mass}
\ee

\subsubsection*{Mass Bound}

We find that the mass term defined in Eq.\,\eqref{eq:Lm} must satisfy the condition
\be
m^2\geq - \frac{(d-1)^2}{4L^2}   
\label{eq:mass_bound}
\ee
to avoid tachyon-like instability of the theory. 
This can be, for example, obtained by putting the equation of motion Eq.\,\eqref{eq:D_mass} in Schr\"odinger form, by going to conformal coordinates $z$, and performing the field redefinition $\Phi= e^{-\frac{d-1}{2L}z} \Psi$. The equation of motion then takes a flat space form, {$\partial^2_z\Psi + \square^{(d)}\Psi -(m^2+\frac{(d-1)^2}{4L^2})\Psi=0$}. In analogy to flat space, requiring the absence of violation of causality implies 
Eq.\,\eqref{eq:mass_bound}. 
The same conclusion can be obtained by inspection of the Feynman propagator in position space, see Eq.\,\eqref{eq:G_pos}. {The negative lower bound in Eq.~(\ref{eq:mass_bound}) is somewhat similar to the AdS Breitenlhoner-Freedman bound, even though our argument is analogous to a flat space one.}

\subsection{The free 2-pt correlators }

It is convenient to work in position-momentum space $(p^\mu,\y)$ to determine the 2-pt functions of the scalar field. We introduce the Fourier transformed 2-pt functions as $G(x,x',r,r')=\int\frac{d^d p}{(2\pi)^d}G(\y,\y';p)e^{ip_\mu (x-x')^\mu}$.
In position-momentum space the equation of motion of the propagator is given by 
\be
{\cal D }_\y G(\y,\y';p)= -i \frac{L^{d}}{\y^{d}} \delta(\y-\y^\prime)  \,, 
\label{eq:EOM_G}
\ee
where the wave operator is given in Eq.~\eqref{eq:D_mass},
and we will introduce
\be
\Delta_p=
\sqrt{L^2 (p^2+m^2)+\frac{ (d-1)^2}{4} -i\epsilon} \,, \quad \gamma_p=\sqrt{-L^2 (p^2+m^2) - \frac{(d-1)^2}{4}+i\epsilon}\,, 
\ee
with $\gamma_p= i\Delta_p$ if  $\epsilon>0$.

\subsubsection{Modes}

The solutions to ${\cal D}_r\Phi = 0$ are 
\be
\y^{\frac{1-d}{2} \pm \Delta_p}\,.
\ee
We can see that the quantity
 $\lambda=\sqrt{\frac{(d-1)^2}{4L^2}+m^2}$ is a threshold: solutions with {$-p^2= p_0^2-\vec p^{\,2}>\lambda^2$} are oscillating, while solutions with $-p^2<\lambda^2$ are exponential.\,\footnote{This behavior has been pointed out in $d=3$ in a number of papers, see e.g. \cite{Asrat:2017tzd,Giribet:2017imm,Chakraborty:2020yka}. Our interpretation  is standard:  the bulk QFT has a Hilbert space with 
gapped continuous spectrum. In analogy with AdS/CFT, these normalizable modes may also be identified as the states of the putative boundary  theory (see e.g. \cite{Balasubramanian:1999ri,Meltzer:2020qbr}.)}

 Upon suitable normalization, the oscillating solutions form a set of  modes $\{f_+,f_-\}$ defined as
 {
 \be
  f_{\pm}(\y,p) = \sqrt{\frac{L}{2\gamma_p}} \left(\frac{\y}{L}\right)^{ \frac{1-d}{2} \pm i \gamma_p}\,.
 \ee
 }
This set of modes satisfies orthogonality under $\int_0^\infty d\y \sqrt{g}\frac{\L^2}{\y^2} $ , that is
{
\begin{align}
&\int_0^\infty d\y \frac{\y^{d-2}}{\L^{d-2}} f_{\pm}(\y,p) f^\ast_{\pm}(\y,p') =\\
&\frac{\L^2}{\sqrt{2\gamma_p}\sqrt{2\gamma_{p'}}} \int_0^\infty \frac{d\y}{\y} e^{i(\gamma_p-\gamma_{p'})\log(\y/L)} =\frac{ \pi \L^2 }{\gamma_p}\delta(\gamma_p-\gamma_{p^\prime})= 2\pi\delta(p^2-p'^2)   \,. \nonumber
\end{align}
}
The set of modes also satisfies a completeness relation obtained by summing over all the allowed values of $p^2$, 
{
\be
\int^{-\lambda^2}_{-\infty} dp^2  f_{\pm}(\y,p) f^*_{\pm}(\y^\prime,p) = \frac{1}{L} \left( \frac{\y\y^\prime}{L^2} \right )^{\frac{1-d}{2}} 
 \int^\infty_0 d\gamma \, e^{i\gamma \log(\y/\y^\prime)}=  \pi  \left( \frac{\y}{L} \right)^{2-d} \delta(\y-\y^\prime)  \,. 
\ee
}
The $\frac{1}{\sqrt{2\gamma_p}}$  normalization  factor of the modes is key for the completeness relation to work out.

\subsubsection{The Wightman propagator}

The Wightman propagator is the free non-ordered 2-pt function $W(x,x')=\langle \Phi(x-i\epsilon) \Phi(x'+i\epsilon)\rangle$ \cite{Streater:1989vi}. It can be thought of as the building block for the other 2-pt functions. 

The Wightman propagator can be obtained as the sum of normalizable modes of posi\-tive energy.  Here we have to sum over both the $f_+$ and $f_-$ modes, which are inequivalent modes, $W\propto\sum_{i=\pm}f_i(r)f_i^*(r')$. 
We obtain 
\be
W(\y,\y';p)=
\frac{ L }{2 \gamma_p}  
\left(\frac{\y \y'}{L^2}\right)^{\frac{1-d}{2}} 
\left(\left(\frac{\y}{\y^\prime}\right)^{i \gamma_p}+\left(\frac{\y^\prime}{\y}\right)^{i\gamma_p}\right)
\theta\left(p^0 - \sqrt{\lambda^2 + {\bm p}^2}\right) 
\ee
{where $\theta(x)$ is the Heaviside function. 

We can see that the symmetries of the background are reflected in the form of the Wightman propagator. 
Firstly,  the Wightman propagator has scaling dimension $\Delta=2\Delta_\Phi=d-1$, it is thus consistent with the scaling obtained in Eq.\,\eqref{eq:Delta_Phi} by requiring invariance of the matter action under dilatation Eq.\,\eqref{eq:DSLD}.  
Second, the Wightman  function is invariant under the inversion symmetry: 
\be
S[W(\y,\y';p)]= \frac{L^{2d-2}}{(\y\y')^{d-1}}W\left(\frac{L^2}{\y},\frac{L^2}{\y'};p\right)=
W(\y,\y';p) \,.
\ee
This symmetry is pictured in Fig.~\ref{fig:inversion}.

Apart from these symmetries,  we can verify that the Wightman function is Hermitian, $W(\y^\prime,\y)^\ast = W(\y,\y^\prime)$, which is always true and serves simply as a sanity check of our calculations.

\begin{figure}[t]
\centering
\includegraphics[trim={0cm 2cm 2cm 1cm},clip,width=0.7\textwidth]{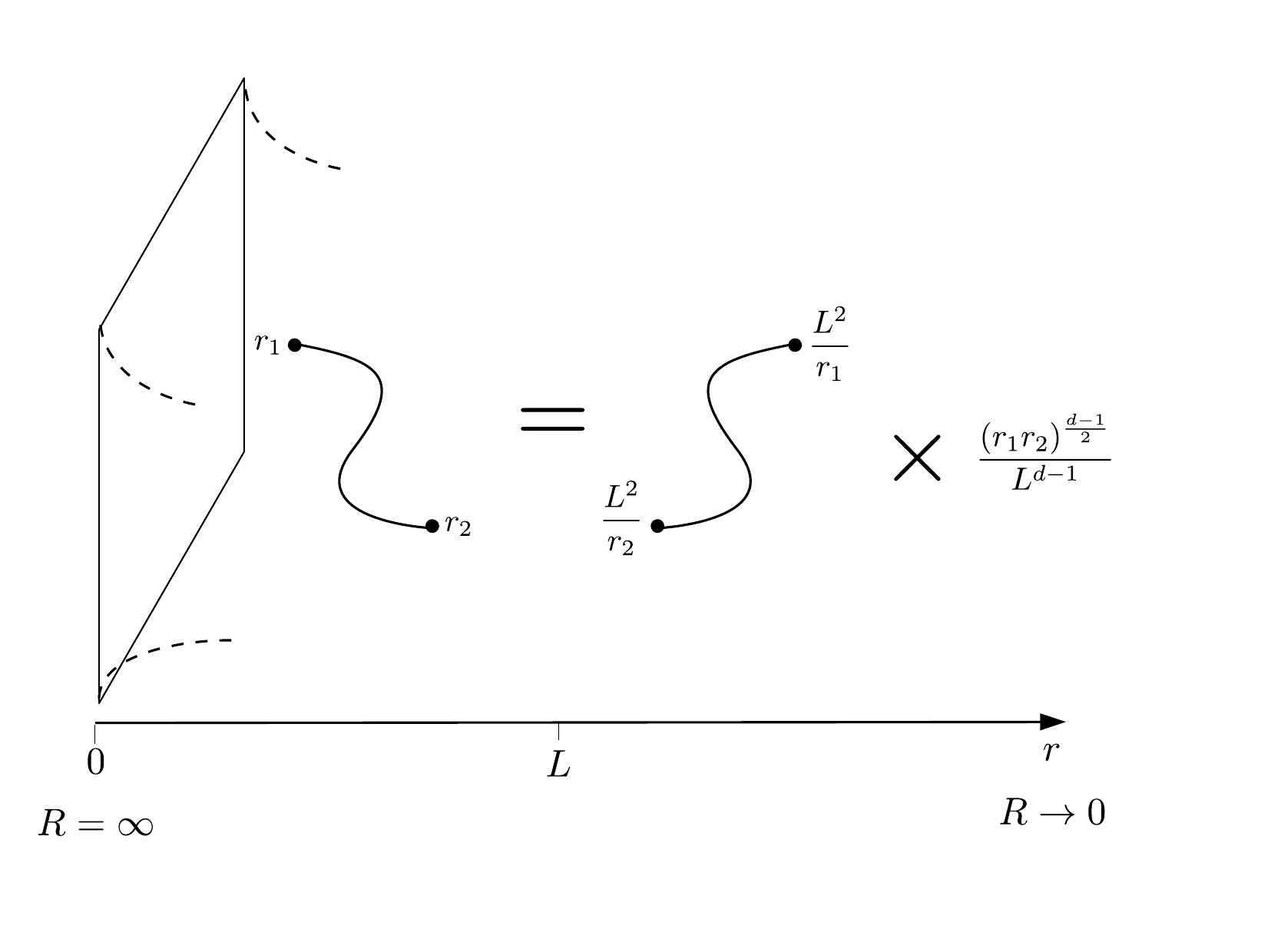}
\caption{\it Action of the $S$ symmetry on a 2-pt correlator in the linear dilaton background.   }
\label{fig:inversion}
\end{figure}

\subsubsection{The Feynman propagator}

The Feynman propagator is defined by 
\be
G(x,x') = W(x,x')\theta(t-t') + W(x',x)\theta(t'-t) \,.
\ee
In position-momentum space we  find
\be
G(\y,\y';p)= - \frac{i L }{2 \Delta_p}  
\left(\frac{\y \y'}{L^2}\right)^{\frac{1-d}{2}} 
\left(  \frac{\y_<}{\y_> } \right)^{\Delta_p} \,,
\label{eq:GLD}
\ee
{where $r_{<}=\textrm{min}(r,r')$, $r_{>}=\textrm{max}(r,r')$}. { Independently from the above derivation, $G$  also corresponds to the solution of the equation of motion Eq.\,\eqref{eq:EOM_G}. }

It turns out that the LD propagator can be expressed in terms of a flat space propagator. 
Going to conformal coordinates for convenience and defining $G(\y,\y';p)\equiv\hat G(z,z';p)$, we find
\be
\hat G(z,z';p) = e^{\frac{d-1}{2}\frac{z+z'}{L}} \hat G_{\rm flat}\left(z,z';p;m^2+\frac{(d-1)^2}{4L^2}\right)  \,, \label{eq:G_LD_G_flat_momentum}
\ee
where $\hat G_{\rm flat}$ is the  propagator of a scalar in $d+1$ dimensions with squared mass $M^2=m^2+\frac{(d-1)^2}{4L^2}$, here expressed in momentum-position space $(p_\mu,z)$.  
The $\hat G_{\rm flat}$ propagator  in \eqref{eq:G_LD_G_flat_momentum} is given by
\begin{align}
\hat G_{\rm flat }&\left( x,x';M^2\right) = \int \frac{d^Dp}{(2\pi)^D} e^{i p_N(x-x')^N} \frac{i}{-p_Np^N-M^2+i\epsilon}  
\\ \nonumber
=  \int & \frac{d^dp}{(2\pi)^d} e^{i p_\mu(x-x')^\mu} \frac{e^{  i\sqrt{-p^2-M^2+i\epsilon} |z-z' | }}{2\sqrt{-p^2-M^2+i\epsilon}} 
 \equiv \int \frac{d^dp}{(2\pi)^d}  e^{i p_\mu(x-x')^\mu} \hat G_{\rm flat} (z,z';p;M^2)
  \,.
\end{align}
Having identified $\hat G_{\rm flat} (z,z';p;M^2)$ as the integrand of the bottom left term, one can easily verify the relation~\eqref{eq:G_LD_G_flat_momentum}. 
Equivalently,  in position space with coordinates $(x^\mu,r)$ and defining  $G_{\rm flat}\left(
x,x';  M^2\right) \equiv \hat  G_{\rm flat}\left(
x,x' ; M^2\right)|_{z=L\log\left(\frac{L}{r}\right)} $, 
the relation is
\be
G\left(x,x'\right)= \left(\frac{\y \y'}{L^2}\right)^{\frac{1-d}{2}} G_{\rm flat}\left(
x,x' ; m^2+\frac{(d-1)^2}{4L^2}
\right) \,.  \label{eq:G_pos}
\ee

By considering a timelike interval with endpoints with identical $\y$, we can notice that the propagator would blow up if $m^2<-\frac{(d-1)^2}{4L^2}$, indicating violation of causality. This proves the mass bound Eq.\,\eqref{eq:mass_bound}.

\subsection{Effective field theory }

Having understood the symmetries of the free theory, we can build an interacting theory by following the principles of effective field theory. 
We will write local operators taking schematically the form ${\cal L}_{\rm eff} \supset F_{m,n}(\y)\partial^{2m} \Phi^n$. The dilatation and the inversion symmetries completely constrain the $F$ function. 

\subsubsection{Bilinear operators}

The bilinear operators of the theory can be built using powers of the following D'Alembertian, \be \widetilde \square_\y \equiv \frac{\y^2}{L^2}\square_\y \,.\ee 
{The quantity $\widetilde\square_r\Phi(r)$} has the same  scaling dimension as $\Phi$, and furthermore transforms as $\Phi$ under the inversion operation: 
\be
S\left[\widetilde \square_\y\Phi(\y)\right] =  \frac{L^{d-1}}{\y^{d-1}} 
 \widetilde \square_{\frac{L^2}{\y}}\Phi\left(\frac{L^2}{\y}\right) =
 \widetilde\square_{\y}\left[\Phi\left(\frac{L^2}{\y}\right)\frac{L^{d-1}}{\y^{d-1}}\right] = \widetilde \square_\y \hat \Phi(\y)  \,.
\ee
We can therefore write the complete bilinear action as 
\be
{\cal L}^{\rm bil}_{\rm LD}= -\frac{1}{2}\int d^dx\, d\y \frac{\y^d}{L^d}\Phi \Pi[\widetilde \square_r]\Phi  \,,
\ee
with the self-energy $\Pi$ expanded as
\be
\Pi[\widetilde \square_r] = \sum^\infty_{n=0} \frac{a_n}{\Lambda^{2n-2}} (\widetilde \square_r)^n\,. 
\ee
The $n=1$ term is the  kinetic term, with $a_1 = 1$ for canonical normalization. The $n=0$ term is the mass term, with {$a_0=\frac{m^2(r)}{\Lambda^2}$.} 

\subsubsection{Interactions}

Interaction terms are similarly constrained by the dilatation and inversion symmetries. For example, for a nonderivative interaction $\Phi^n$ we find
\be
{\cal L}^{\rm int}_{\rm LD}\supset -\frac{\gamma_n}{n!}\int d^dx\, d\y \left(\frac{\y}{\L}\right)^{n\frac{d-1}{2}-1}\Phi^n\,.
\ee
{As in the case of the mass term, we could also write this interaction covariantly 
 with a dilaton-dependent coupling $\gamma_n(\phi)\equiv (kLe^{\bar\phi} )^{d+1-n\frac{d-1}{2}  }\gamma_n$.} 
We can also check that  the derivative interaction
\be
{\cal L}^{\rm int}_{\rm LD} \supset -\frac{1}{n!}\int d^dx\, d\y \left(\frac{\y}{\L}\right)^{(n+2)\frac{d-1}{2}+1}\Phi^n \partial_M\Phi \partial^M \Phi \, 
\ee
is invariant. The invariance under inversion is not straightforward. It happens due to cancellations,  upon integration by part of a $\Phi' \Phi^{n+1}$ term which appears when applying the symmetry. 

\subsubsection{Structure of  correlators}

\label{se:LD_correlator_gen}

Consider a diagram contributing to an $l$-pt correlator
$\langle\Phi(r_1)\ldots\Phi(r_l)\rangle_{{\rm LD}_{d+1}}\equiv{\cal M}^{(l)}_{{\rm LD}_{d+1}}$. In position space, the diagram is built using the above interactions together with the position space Feynman propagator of Eq.\,\eqref{eq:G_pos}. Most  $\frac{\y}{\L}$ powers simplify inside the diagram, the remaining ones combine such that the interior of the diagram amounts to an amplitude in flat space  with  $(x^\mu,\L\log(\frac{L}{\y}))$ coordinates (or equivalently $(x^\mu,z)$ in conformal coordinates). 
The only remaining $\frac{\y}{L}$ factors are those associated with the endpoints of the external propagators. 

The above facts put together imply  that any Feynman diagram of our EFT  takes the form of a \textit{flat space} diagram with each external leg  ending at $\y_i$  multiplied by a $\left(\frac{\y_i}{\L}\right)^{\frac{1-d}{2}}$ factor, {i.e.}
{
\be
{\cal M}^{(l)}_{{\rm LD}_{d+1}}(r_1,\dots,r_l)={\cal M}^{(l)}_{{\rm flat}_{d+1}}(r_1,\dots, r_l) \prod_{i=1}^l\left(\frac{\y_i}{\L}\right)^{\frac{1-d}{2}}\,. \label{eq:M_LD_flat}
\ee
}
 The emergence of a flat space amplitude is a consequence of the symmetries we have imposed to build the EFT.
The ${\cal M}_{{\rm flat}_{d+1}}(r_i) $ correlator is invariant under the dilatation of the $\y_i$ coordinates (or equivalently under translation in the conformal coordinates $z_i$). The scaling behavior of ${\cal M}_{{\rm LD}_{d+1}}(\y_i)$ 
is thus simply set by the overall powers of $\y$ in the right-hand side of \eqref{eq:M_LD_flat}. This  ensures that  the   {${\cal M}_{{\rm LD}_{d+1}}(\y_i) $} diagrams scale as $l\frac{d-1}{2}$, consistent with the scaling dimension 
of a product of $l$ free fields (see definition in section \ref{se:dilatation_field}).

\section{Linear Dilaton Holography} 
\label{se:correlators}

We turn to the holographic properties of the linear dilaton spacetime. We place a flat brane in the spacetime and study  the correlators ending on it.\,\footnote{In this paper we  study  holography on a timelike surface.
Another direction could be to study holography  at  the conformal null boundary of LD (with no brane), just like for flat spacetime, see {e.g.}~\cite{Pasterski:2021rjz} and references therein. This is a distinct analysis that we leave for future work. 
}
From these, we deduce  properties that the putative   $d$-dimensional  holographic dual theory, if it exists, should satisfy. 
We also point out that the properties of the holographic correlators can be explored using  techniques analogous to those applied to the wavefunction of the Universe. 

\subsection{Two halves of LD }
\label{se:halves}

We assume the existence of a flat brane ({i.e.}~a domain wall) along the  $r=r_b $ surface. All the fields have boundary conditions on the brane. 
The brane is described by appending the action
\be
\S_{\rm brane} = -\int d^dx \sqrt{\bar g} \left(V_b(\phi)+\Lambda_b\right)
\label{eq:action_brane}
\ee
to the bulk action defined in \eqref{eq:action_bulk}.
The brane carries a localized potential $V_b(\phi)$.  The induced metric determinant in our coordinates is $\sqrt{\bar g}=(r_b/L)^d$. 

We assume the brane potential sets the value of $\phi$ to $\phi=v_b$, and  
$\Lambda_b$ is tuned to $-V_b(v_b)$ so that the total brane tension is zero. 
The dilaton vacuum expectation value (vev) introduces a new physical scale
\be
\eta \equiv k \, e^{\bar v_b} \,. \label{eq:eta_k}
\ee

The solution to the bulk field equations is the same as without a brane, but the parameters of the solution differ. While the brane location $r_b$ may in principle evolve in time (see section \ref{subsec:time_evolution}), the value of $\bar v_b$ does not change as a function of $r_b$, because it is fixed by a brane-localized potential, {i.e.}~$\frac{\partial \bar v_b}{\partial r_b}=0$.  
This fact implies a constraint on  the $c$ parameter,
\be
c = \frac{1}{\eta r_b}\,. \label{eq:c_brane}
\ee
 Other equivalent conditions are given in  section \ref{se:conditions}.
The solution of the field equations with the parameter fixed in Eq.\,\eqref{eq:c_brane}
is
\begin{equation}
ds^2 =  \frac{1}{\eta^2 r_b^2} d\y^2 + \frac{\y^2}{\L^2}\eta_{\mu\nu}dx^\mu dx^\nu \,,  \qquad \bar \phi(r) = \log \left( \frac{\y_b}{\y} \right) + \bar v_b  \,. 
\end{equation}
}

\begin{figure}[t]
\centering
\includegraphics[trim={0cm 3cm 2cm 1.5cm},clip,width=0.8\textwidth]{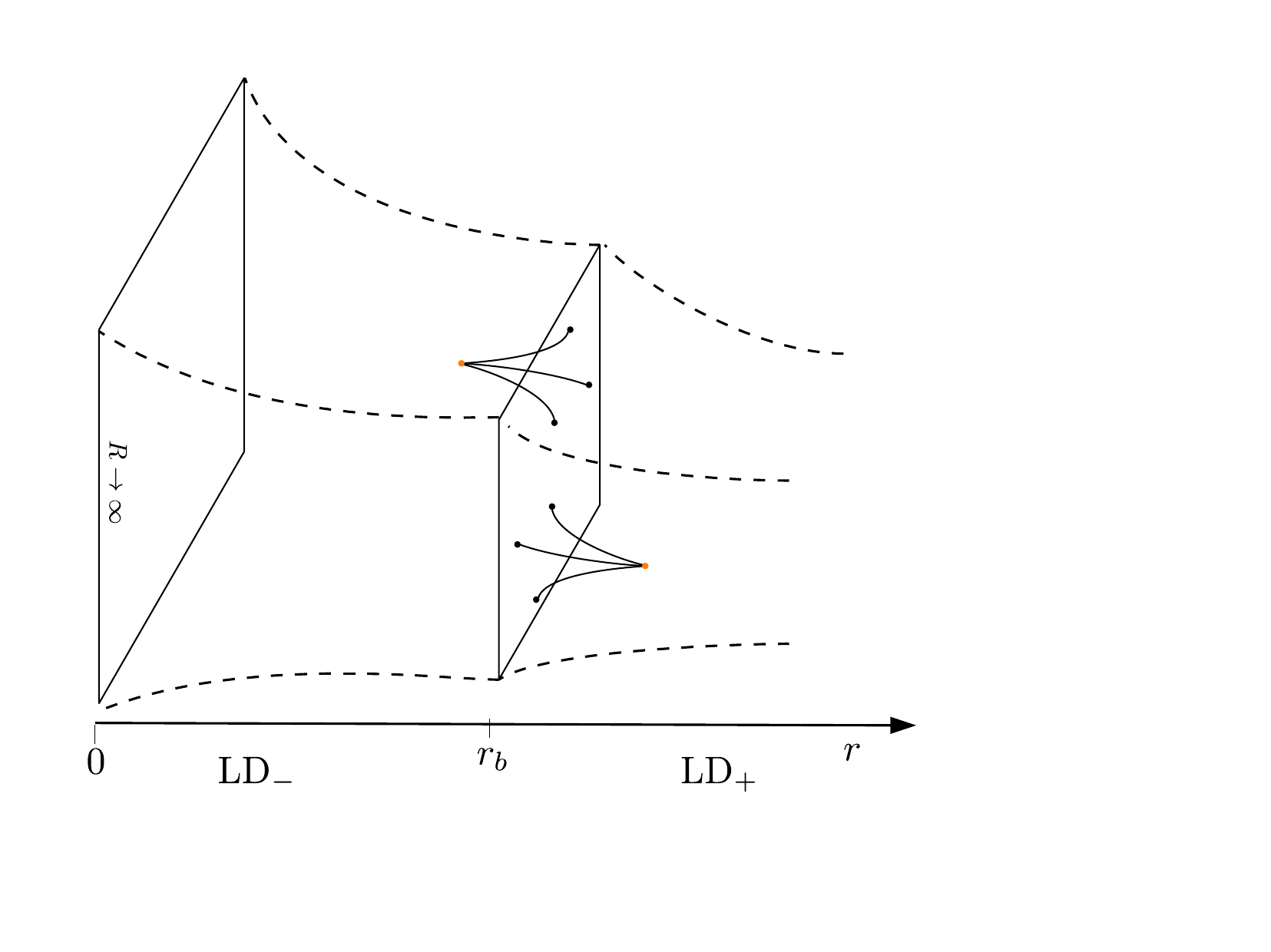}
\caption{\it  The linear dilaton spacetime with a brane.  The {\rm LD}$_\pm$   regions can support different {QFT}s. 
The drawn 3-pt contact diagrams  illustrate brane correlators in each region.        }
\label{fig:LD_brane}
\end{figure}

The brane  splits the spacetime into two regions, as illustrated in Fig.\,\ref{fig:LD_brane}. 
We refer to the $r\in[0,r_b]$ region as LD$_-$, 
and to the $r\in[r_b,\infty]$ region as LD$_+$. 
The two  regions are fundamentally different: LD$_-$ contains the singularity while LD$_+$ does not. Our goal is to study the quantum  fields living on the  LD$_\pm$ regions from the viewpoint of the brane.

\subsubsection{Preliminary observations}

Properties of the QFT differ between LD$_-$ and LD$_+$ regions. As will be shown below, in the LD$_-$ region the spectrum in $d$-momentum space  features isolated discrete modes in addition to a gapped continuum.  These \isolated modes are absent in region LD$_+$. A qualitative way to understand this  phenomenon is that the quantum fluctuations tend to be repelled from the curvature singularity, so that in the LD$_-$ region there is a notion of confinement between the  brane and the singularity.

Various truncations of the LD background have been considered in the literature~\cite{Antoniadis:2001sw, Antoniadis:2011qw, Antoniadis:2021ilm,Megias:2021mgj}, sometimes involving two branes. Our analysis provides a simple overview of the regions and makes it clear that \isolated modes exist even with a single brane. \footnote{Solutions for $D=5$ in the presence of branes have been studied in Refs.~\cite{Antoniadis:2011qw, Antoniadis:2021ilm,Megias:2021mgj}. 
Consider $A(z)= \sigma z$, where the parameter $\sigma \propto 1/L$ is such that ${\rm sign}(\sigma)={\rm sign} (a)$, see Eq.~\eqref{eq:Aphi_z}.  The $\sigma < 0$ case was studied in \cite{Antoniadis:2011qw, Antoniadis:2021ilm}  and
 the  $\sigma > 0$ case in \cite{Megias:2021mgj}. In our analysis, these cases map respectively to LD$_+$ and LD$_-$, which makes it clear why there is low-energy gravity in the latter, and not in the former. }

As will be clear below, the \isolated modes of the gravity sector in the LD$_-$ region are a massless graviton and a massive scalar, usually called radion~\cite{Megias:2021mgj}. 
The existence of the massive radion  mode ensures that the brane-singularity distance is stable in the one-brane setup. 

 The existence of the isolated massless graviton in the LD$_-$ region means that {e.g.}~one can exploit the $D=5$ case  to build models featuring Einstein-like 4d gravity at low energy. This phenomenological line has been followed to some extent in~\cite{Megias:2021mgj}. 
In contrast, in the LD$_+$ region there is no graviton massless mode. Hence at energies below the graviton mass gap, gravity can be integrated out.  In the LD$_+$ region we thus recover the familiar LST property that gravity decouples at low-energy \cite{Aharony:1999ks,Kutasov:2001uf}.

Our focus in this work is \textit{not}  on the \isolated modes of the LD$_-$ region, but rather on the theory as seen from the brane. That is, from the viewpoint of the partition function, we probe the theory by placing sources on the brane. Integrating out the bulk modes  defines the brane quantum effective action, which generates the brane correlators. Unlike an EFT for \isolated modes,  this holographic viewpoint applies for both LD$_-$ and LD$_+$ regions. 
Integrating out either the  LD$_+$ or LD$_-$ region of the bulk gives rise to two inequivalent holographic effective theories described by the quantum effective actions $\Gamma_+$ and $\Gamma_-$.

Throughout this section we are not interested in the $r_b$ dependence of the correlators. We then simplify the expressions by assuming {the brane located at}
\be
r_b\equiv \frac{1}{\eta} \,.\label{eq:rb_hyp}
\ee
For the correlators we compute in this section, assumption
\eqref{eq:rb_hyp} 
is equivalent to stating that the correlators $\hat G \equiv G|_{r_b=\eta^{-1}}$ are functions of the rescaled coordinates 
$(\hat x_\mu , \hat r)$ with \be  \hat x_\mu = x_\mu {\eta r_b}\,, \qquad \hat r = \frac{r} {\eta r_b} \,,
\label{eq:redef}
\ee
({i.e.}~$\hat p_\mu = p_\mu ({\eta r_b})^{-1}$). The general case $G(x_\mu, r)$ can be recovered by plugging \eqref{eq:redef} into the $\hat G$ correlators, $G(x_\mu, r)=\hat G (\hat x_\mu, \hat r)$.

\subsection{Fields and Propagators }

We study the quantum fluctuations  of a matter scalar field $\Phi$ and of the bulk metric in the LD$_\pm$ regions. 
We assume that the propagators are regular at $\y\to0$ and $\y\to\infty$, just like in the LD spacetime without brane.  
We assume Neumann boundary condition for the matter scalar. 
For the metric fluctuation our focus is the tensor component $g_{\mu\nu}$, {i.e.}~the $d$-dimensional graviton, 
which has automatically Neumann boundary condition on the brane.

\subsubsection{Matter Scalar}

We consider the matter scalar field $\Phi$ living on the LD background in the presence of the brane. The action and properties of this field in the full LD background have been studied in Sec.~\ref{se:fields}. We denote its bulk action as ${\cal S}^{\rm bulk}_\Phi$. The full action describing the effect of the brane on $\Phi$ is
\be
{\cal S}_\Phi = {\cal S}^{\rm bulk}_\Phi  -  \frac{1}{2}\int_{\rm brane} d^dx \frac{r_b^d}{L^d} \Phi{\cal B}[\square^{(d)}] \Phi \label{eq:SPhibrane}\,.
\ee
The surface term, which contains the function ${\cal B}$,  contains brane-localized operators which are bilinear in  $\Phi$ and depend on the flat-space $d$-dimensional d'Alembertian.

Upon Fourier transform along the flat slices, the bulk action gives rise to the bulk  wave operator ${\cal D}$  given in \eqref{eq:D_mass}. 
The corresponding bulk equation of motion is given in \eqref{eq:EOM_G}.
  The Neumann boundary conditions on the Green's functions of $\Phi$ are  
\be
(\partial_\y\pm{\cal B})G(\y,\y^\prime;p)|_{\y=\y_b} =0 \,,\; \forall\, \y, \y^\prime \in {\rm LD}_\mp \,.
\ee
With this boundary condition, the Feynman  propagator $G_\Phi^\pm$ in regions LD$_\pm$ is found to be
\be
G^\pm_\Phi(\y,\y';p)= \frac{i\L}{2\Delta_p}  
\left(\frac{\y \y'}{L^2}\right)^{\frac{1-d}{2}} 
\left(b_p \left(\frac{L^2}{\y \y'}\right)^{\pm\Delta_p} - \left( \frac{\y_<}{\y_>}\right)^{\Delta_p} \right)  \,, 
\label{eq:G_N}
\ee
with 
\be
b_p=\frac{ 1-d\pm2\Delta_p \mp 2 \y_b{\cal B} }{1-d\mp 2\Delta_p \mp 2 \y_b{\cal B}} \left( \frac{\y_b}{L} \right)^{\pm2\Delta_p} \,,
\ee
{where $\mathcal B\equiv\mathcal B(-p^2)$.}

\subsubsection{Graviton}

We define the graviton as the traceless fluctuation $h_{\mu\nu}$ of the metric 
\begin{equation}
ds^2 =  d\y^2+ \frac{\y^2}{\L^2}\left(\eta_{\mu\nu} + h_{\mu\nu}(x,\y)\right) dx^\mu dx^\nu \,, 
\end{equation}
{where we have used Eq.~(\ref{eq:rb_hyp}).}
In order to determine the graviton spectrum, it is enough to focus on the transverse  component of $h_{\mu\nu}$, 
$h^{\perp}_{\mu\nu}= P^{\perp}_{\mu\rho}P^{\perp}_{\nu\sigma} h^{\rho\sigma} -\frac{1}{4}\eta_{\mu\nu} P_{\rho \sigma}^\perp h^{\rho\sigma} $, with $P^{\perp}_{\alpha\beta}=\eta_{\alpha\beta}-\frac{\partial_\alpha \partial_\beta}{\square}$. 
The $h^{\perp}_{\mu\nu}$ component contributes to the graviton propagator as 
\be
G^h_{\mu\nu,\rho\sigma}(x,x') = G_h(x,x') 
\mathbb{I}_{\mu\nu,\rho\sigma}
+ \ldots   \,,
\label{eq:Gh}
\ee
with 
$\mathbb{I}_{\mu\nu,\rho\sigma}=
 \frac{1}{2}\left(\eta_{\mu\rho}\eta_{\nu\sigma} + \eta_{\mu\sigma} \eta_{\nu\rho} \right) - \frac{1}{d}\eta_{\mu\nu}\eta_{\rho\sigma}
$ the identity on the space of $d$-dimensional traceless symmetric tensors.
The  equation of motion of the reduced graviton propagator $G_h$
is  simply expressed with the scalar wave operator 
\be
\square_r G_h(x,x')= i\frac{\L^d}{\y^d}\delta^{d+1}(x-x')\,,
\ee
with boundary condition $\partial_r G_h|_{r=r_b} = 0$ on the brane.  

The reduced bulk Feynman  propagator in the LD$_\pm$ regions is related to the scalar one. We find
\be
G^\pm_h(\y,\y';p)= G^\pm_\Phi(\y,\y';p) \bigg|_{m=0, \, {\cal B}=0}\,. 
\ee
Since the graviton propagator can be expressed in terms of the scalar propagator, our focus below is essentially on the scalar sector. We shall drop the $\Phi$ indexes from the scalar quantities, while the graviton case is obtained by specializing to $m\to 0$, ${\cal B}\to 0$.

\subsection{Holographic Correlators}
\label{se:hol_corr}

We now compute the entries of the boundary quantum effective action of the fields, denoted by $\Gamma_\pm[\Phi_b]$, where the $\Phi_b$ argument is the classical field value on the brane. The effective actions $\Gamma_+$ and $\Gamma_-$ are obtained from integrating out, respectively, the LD$_+$ and LD$_-$ regions. They are in principle different. 

The boundary quantum effective action can be conveniently evaluated by expressing the fields in a holographic basis which separates the bulk and brane degrees of freedom. In momentum space it takes the form  $\Phi(p;\y)=\Phi_b(p) K(p;\y)+\Phi_{ D}(p;\y) $, where $\Phi_b(p)$ is the boundary degree of freedom, and $\Phi_{ D}(p;\y)$ is the Dirichlet component satisfying $\Phi_{D}(p;r_b)=0$. 
The $K(p;r)$ function, {which satisfies the condition $K(p;r_b)=1$ for any $p$,} can be chosen to be the amputated Neumann brane-to-bulk propagator. With this choice the quadratic action is diagonal in the $(\Phi_b,\Phi_{D})$ basis \cite{Fichet:2021xfn}.  
There is a holographic decomposition for each of the LD$_\pm$ regions, with distinct  brane-to-bulk propagators $K_\pm$, that are given further below.

The effective action $\Gamma_\pm[\Phi_b]$ contains the  1PI pieces of the boundary correlators, that can be extracted  by taking derivatives in $\Phi_b$. Notice that  the notion of 1PI is meant with respect to boundary-to-boundary lines. 
Our focus is the tree-level entries of $\Gamma_\pm[\Phi_b]$, sometimes referred to as holographic self-energy and vertices.

%Our focus is tree-level 1PI entries, that we denote ${\cal C}^{l}$ 

The tree-level bilinear entry of $\Gamma_\pm[\Phi_b]$, {i.e.}~the \textit{holographic self-energy}, is given by 
\be
{\cal C}_\pm^{(2)}(p)\equiv \frac{1}{\sqrt{\bar g}}\frac{\delta^2 \Gamma_\pm[\Phi_b]}{\delta \Phi_b(p)\delta \Phi_b(-p)} = \mp \partial_\y K_\pm(p,\y)|_{\y=\y_b} =\frac{1}{\sqrt{\bar g}}  \frac{1}{i G^\pm(p,\y_b,\y_b)} \,. \label{eq:HSEdef}
\ee 
 The last equality is a general property of the holographic decomposition which similarly holds for any background~\cite{Fichet:2021xfn}.

The higher point terms of $\Gamma[\Phi_b]$, {i.e.}~the~\textit{holographic vertices}, amount to bulk  diagrams built from Dirichlet propagators, with the $K_\pm(p_j,\y_j)$ in external legs. 
They take the general form
\be
{\cal C}_\pm^{(l>2)}(p_i)= 
\left(\prod_{j=1}^l\int dr_j K_\pm(p_j,\y_j)\right) {\cal A}_\pm^{(l>2)}(p_i,\y_i) \label{eq:C_gen}
\ee
where ${\cal A}_\pm^{(l>2)}(p_i,\y_i)$ is an  $l$-pt bulk subdiagram made from Dirichlet lines. 
We use the notation ${\cal C}^{(l>2)}(p_i)\equiv {\cal C}^{(l>2)}(p_1,\ldots,p_l)$ and similarly for ${\cal A}$ and related quantities.

\subsubsection{Propagators}

\label{se:prop_hol}

The relevant  propagagors in the LD$_\pm$ regions are as follows. 
The (amputated) brane-to-bulk propagators are
\be
K_\pm(p;\y)=\left(\frac{\y}{\y_b}\right)^{\frac{1-d}{2}\mp\Delta_p}\,.
\ee
The Dirichlet propagators that appear in internal bulk lines
are 
\be
G^\pm_{D}(\y,\y';p)= \frac{i\L}{2\Delta_p}  
\left(\frac{\y \y'}{L^2}\right)^{\frac{1-d}{2}} 
\left( \left(\frac{r_b^2}{\y \y'}\right)^{\pm\Delta_p} - \left( \frac{\y_<}{\y_>}\right)^{\Delta_p} \right)  \,. 
\ee

The holographic basis provides a decomposition of the Neumann propagator into a Dirichlet propagator plus a boundary contribution describing the propagation of the boundary degree of freedom. This  Neumann-Dirichlet identity in the LD$_\pm$ regions is 
\be
G^\pm(\y,\y';p) = G^\pm_{D}(\y,\y';p) + K_\pm(p,\y)G^\pm(\y_b,\y_b;p)K^\pm(p,\y')\,. 
\label{eq:ND_rel}
\ee

\subsubsection{Dual theory}

Finally,  we can always interpret $\Gamma[\Phi_b]$ as the generating functional of an unknown $d$-dimensional theory probed by the $\Phi_b$ source,
\be
\Gamma[\Phi_b] \equiv \left\langle e^{i\int dx^d \Phi_b {\cal O}} \right\rangle_{\rm  dual~(unknown) }
\ee
with $\cal O$ an operator from the dual theory. 
The ${\cal C}^{(l)}(p_i)$ are then interpreted as the connected correlators of this putative dual theory. 

The dual theory is unknown in the sense that we do not know the $d$-dimensional fundamental action that describes it.
Whether this dual description actually exists is a highly nontrivial question. 
In this work we  infer from  $\Gamma[\Phi_b]$ 
some generic features that the dual theory should possess. We emphasize that all  our analysis remains relevant independently from the actual existence of the dual theory.

\subsection{Two-point Correlators and Spectrum}

\label{se:2-pt_corr}

The scalar brane-to-brane propagators in the LD$_\pm$ regions are
\be
G^\pm_{b}(p^2)\equiv G^\pm(\y_b,\y_b;p)= -i\frac{\L^d}{r_b^{d-1}} \frac{1}{\Delta_p\pm\frac{d-1}{2}+\y_b {\cal B}}\,, \label{eq:GbbScalarLD}
\ee
From the viewpoint of the brane, these are 2-pt functions  in $d$-dimensional flat space. They thus  have a Källén-Lehmann representation from which we can  extract a spectral function. In momentum space the spectral function is given by
{
\be
\rho^\pm(\mu) = \frac{1}{\pi }{\rm Im}[i G^\pm_b(-\mu^2)]\,.
\ee
}

For convenience we fix the brane operator to be a mass term, i.e.~${\cal B}\equiv \B / r_b$ {where $\bar{\mathcal B}$ is dimensionless}. 
 We find 
\begin{align}
\rho^\pm(\mu) = & \frac{1}{\pi}\frac{\L^d}{r_b^{d-1}} \frac{\sqrt{(\mu^2-m^2)\L^2 -\frac{(d-1)^2}{4}}}{ (\mu^2-m^2)\L^2  -(d-1 \pm \B)  \B } \theta\left(\mu^2-\frac{(d-1)^2}{4\L^2} +m^2 \right) 
\nn \\ & + \delta_{\pm, -}\,\frac{(d-1-2\B)\L^{d-2}}{r_b^{d-1}} \theta\left( \frac{d-1}{2}-\B\right)\delta(\mu^2-m^2_0) \,, 
\label{eq:rhoPhiLD}
\end{align}
with the \isolated mode mass given by {$ m^2_0 = m^2+\frac{(d-1 - \B) \B }{\L^{2}} $}.

\subsubsection{The \isolated mode}

From Eq.\,\eqref{eq:rhoPhiLD} we see that only the region LD$_-$ features an \isolated mode. This can be understood as an effect of confinement between the brane and the singularity. This illustrates the idea that the singularity repels the scalar fluctuations. 

The condition on $\B$ from the Heaviside function on the second line in Eq.\,\eqref{eq:rhoPhiLD} appears because if $ \B \geq \frac{d-1}{2}$ the residue of the pole vanishes. This mechanism automatically prevents the residue to become negative, {i.e.}~it prevents the presence of a ghost in the spectrum. 

We  notice that a massless mode can be obtained by tuning $\B$,
 but only if the bulk mass satisfies  $m^2\geq -\frac{(d-1)^2}{4\L^2}$. Interestingly, this is the same bound as the one obtained in section \ref{se:Mass}, arising here  from a subtler criterion: requiring that a massless \isolated mode \textit{can} exist
 in the parameter space.\,\footnote{
In analogy with a slice of AdS space \cite{Gherghetta:2010cq}, the existence of such a massless scalar mode would be  ne\-cessary in order to  build  supersymmetric multiplets in the low-energy EFT of \isolated modes. 
Aspects of supersymmetry in the LD spacetime have  been investigated in \cite{Antoniadis:2021ilm}.
} 

The graviton case is obtained by setting $\B=0$, $m=0$. Hence the
\isolated graviton mode {in LD$_-$} is massless and the continuous component of the graviton spectrum in LD$_+$ and LD$_-$ are exactly equal, {with a mass gap at  $(d-1)/(2L)$.}   

\subsubsection{Holographic correlators}

We turn to the self-energy ${\cal C}_\pm^{(2)}(p^2)$, defined by Eq.\,\eqref{eq:HSEdef}. This self-energy can always be interpreted as the 2-pt correlator of an unknown $d$-dimensional dual theory. 
The spectral distribution of the 2-pt holographic correlator is given by 
\be
\rho^\pm_{{\cal C}}(\mu) = \frac{1}{\pi }{\rm Im}\left[{\cal C}^{(2)}_\pm(-\mu^2)\right] \,.
\ee
The result from either the LD$_-$ or LD$_+$ regions  is 
\be 
\rho^\pm_{{\cal C}}(\mu) =\frac{\gamma_\mu}{\pi r_b}\theta\left(\mu^2-\frac{(d-1)^2}{4\L^2} -m^2 \right)  \,,
\label{eq:rho_cont}
\ee
where we remind the definition {$\gamma_\mu=\sqrt{(\mu^2-m^2)L^2-\frac{(d-1)^2}{4}}$.} 
We see that the spectral functions are equal. No information on the \isolated mode remains in $\rho_{\cal C}^\pm$.  More generally, the ${\cal B}$ term entirely drops because it enters as a purely analytical term, which cannot contribute to the spectral distribution.

We can proceed similarly with the graviton field. The brane-to-brane propagator for the graviton in LD$_\pm$ regions is the same as Eq.~(\ref{eq:GbbScalarLD}) with $m = 0$ and $\mathcal B = 0$, thus the graviton spectral distribution is
\be 
\rho^\pm_{{\cal C}}(\mu) \Big|_{m=0}\,.
\label{eq:rho_cont_h}
\ee

These results show that the putative dual theory emerging from the holography of  either LD$_\pm$ must describe a \textit{gapped continuum}. A similar conclusion is obtained from other fields, see {e.g.}~\cite{Csaki:2018kxb,Megias:2019vdb,Megias:2021mgj,Megias:2021arn} for related results.  
The fact that both regions lead to the same continuum spectrum is nontrivial. It comes from the fact that the 2-pt propagators in LD$_-$ and LD$_+$ only differ by the \isolated mode, which  is reminiscent of the inversion symmetry of the LD spacetime. 
Interestingly enough, a similar spectrum is also obtained from the single-trace deformed CFT$_2$ studied with string techniques, see Refs.\,\cite{Asrat:2017tzd,Giveon:2017nie,Giribet:2017imm}.

\subsection{Higher-point Correlators and Singularities}

We now investigate 3-pt and 4-pt correlators. 
We use the bulk vertices from the EFT obtained in Sec.~\ref{se:fields}. These vertices respect the  dilatation symmetry by construction.

\subsubsection{Contact diagrams}    

To warm-up we compute a 3-pt contact diagram induced by the cubic vertex 
\be
{\cal S}_{\rm int} \supset - \int d^d x\, d\y \left(\frac{\y}{\L}\right)^{\frac{3d-5}{2}}\frac{\gamma}{3!}\Phi^3\,.
\ee
The diagram in the LD$_\pm$ region, for example, is given by 
\be -i \gamma \int^{\infty(r_b)}_{r_b(0)} dr \left(\frac{\y}{\L}\right)^{\frac{3d-5}{2}}  K_\pm(r,p_1)K_\pm(r,p_2)K_\pm(r,p_3)\,. \ee
We find identical correlators for LD$_-$ and LD$_+$, 
\begin{align}
 {\cal C}_{\pm}^{(3),\,{\rm con}}(p_i)  = 
       \vcenter{
        \hbox{
        \includegraphics[trim={10cm 7.7cm 12cm 7cm},clip,width=0.13\textwidth]{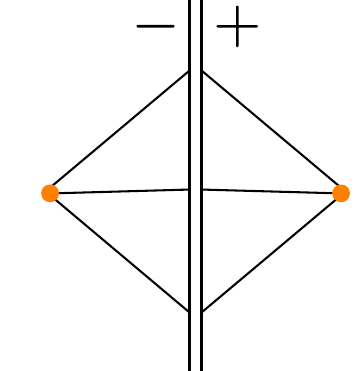}
        }
        }
=
-i L \left(\frac{\y_b}{\L}\right)^{\frac{3}{2}(d-1)} \frac{\gamma}{\Delta_{p_1}+\Delta_{p_2}+\Delta_{p_3}}\,. \label{eq:C3}
\end{align}
We can see that ${\cal C}_{\pm}^{(3),\,{\rm con}}$ has a singularity going as the inverse of the sum of  all scaling dimensions,  $\Delta_{p_1}+\Delta_{p_2}+\Delta_{p_3}\equiv\Delta_T=0$.

Similarly, the $n$-pt contact diagrams from a $\Phi^n$ monomial
are identical in both regions, {${\cal C}_{+}^{(n),\,{\rm con}}={\cal C}_{-}^{(n),\,{\rm con}}$, and are proportional to $1/\Delta_T$ with 
$\Delta_T\equiv \sum_{i=1}^l\Delta_{p_i}$} the sum of all the scaling dimensions. 

\subsubsection{Exchange diagram}  

We compute an $s$-channel 4-pt exchange diagram induced by the cubic vertices.
We denote the exchanged $d$-momentum  by $q_\mu= (p_1+p_2)_\mu$. It appears in the scaling dimension of the propagator, $\Delta_q$. 
We define 
\be
\Delta_T = \sum^4_{i=1}\Delta_{p_i}\,,~\quad \Delta_{ij} = \Delta_{p_i}+\Delta_{p_j}+\Delta_{q}\,. 
\ee

We first compute the exchange diagram with an internal Dirichlet propagator. This is, in the language of Sec.~\ref{se:hol_corr},  a  contribution to the 4-pt holographic vertex.
We find that  this exchange diagram   is
\begin{align}
{\cal C}^{(4),\, {\rm ex}}_\pm (p_i) 
=        \vcenter{
        \hbox{
        \includegraphics[trim={10cm 7.5cm 12cm 6cm},clip,width=0.13\textwidth]{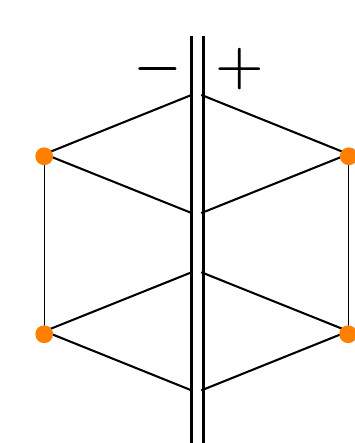}
        }
        }
= i\gamma^2 L^3 \left(\frac{\y_b}{\L}\right)^{2d-2}
\frac{1}{\Delta_{12} \Delta_{34}\Delta_T} 
\, \label{eq:C4LD}
\end{align}
in both regions LD$_\pm$.

In order to compute the exchange diagram with a Neumann internal propagator, we can equivalently re-do the calculation using the propagator from Eq.\,\eqref{eq:G_N}, or add to Eq.\,\eqref{eq:C4LD} the contribution from the  boundary term in \eqref{eq:ND_rel}. 
The Neumann exchange diagram takes the form
\be
{\cal C}^{(4),\, {\rm ex}}_{\pm,N} (p_i)= {\cal C}^{(4),\, {\rm ex}}_\pm (p_i)+ {\cal C}^{(4),\, {\rm ex}}_{\pm, {\partial}} (p_i)
\ee
with the boundary contribution
\be
{\cal C}^{(4),\, {\rm ex}}_{\pm, {\partial}} (p_i) = 
 i\gamma^2 L^3 \left(\frac{\y_b}{\L}\right)^{2(d-1)}
 \frac{1}{\Delta_{12} \Delta_{34}} 
 \frac{1}{  r_b{\cal B} + \Delta_{q} \pm \frac{d-1}{2} }\,.
\label{eq:C4LD_boundary}
\ee

The boundary contribution ${\cal C}^{(4),\, {\rm ex}}_{\pm, {\partial}} $ takes the form of two holographic 3-pt vertices related by a brane-to-brane propagator.   
Accordingly, the ${\cal C}^{(4),\, {\rm ex}}_{-, {\partial}} $  term features the isolated pole with mass $m_0$ that we have already encountered in Sec.~\ref{se:2-pt_corr}. 
The ${\cal C}^{(4),\, {\rm ex}}_{\pm, {\partial}} $ is not  1PI since it features a single  boundary line, unlike  the  ${\cal C}^{(4),\, {\rm ex}}_\pm $ contribution.

We see that the holographic decomposition   naturally separates 
${\cal C}^{(4),\, {\rm ex}}_{\pm,N}$ as  a boundary term and a 1PI term. The boundary term can be built from lower-order 1PI holographic vertices. Hence the nontrivial information is rather encoded into the 4-pt 1PI term ${\cal C}^{(4),\, {\rm ex}}_{\pm}$. Interestingly, it is identical in both regions LD$_\pm$.

The ${\cal C}^{(4),\, {\rm ex}}_\pm (p_i)$ term
features a singularity at $\Delta_T\to 0$. This singularity   is reached for physical values of the external momenta. It occurs when   the four scaling dimensions simultaneously vanish, 
{\be \Delta_{p_1,p_2,p_3, p_4}\to 0
\label{eq:sim_threshold}
\,.\ee
}
The vanishing of a given $\Delta_{p_i}$ corresponds to the threshold value $-p_i^2=m^2+\frac{(d-1)^2}{4\L^2}$ beyond which the 2-pt line (here the $K(p_i,\y_i)$) develops an imaginary part.
This is, physically, the kinematic threshold of the gapped continuum  found in Eq.\,\eqref{eq:rho_cont}.  
The $\Delta_T=0$ singularity thus occurs 
at the configuration of \textit{simultaneous kinematic thresholds}, Eq.\,\eqref{eq:sim_threshold}.

Since the $\Delta_T\to 0$ singularity has a physical meaning, the associated residue may also have one. We find that the residue of the $1/\Delta_T$ pole is 
\be
L \left(\frac{\y_b}{\L}\right)^{2(d-1)}\frac{i \gamma^2 }{q^2+m^2 +\frac{(d-1)^2}{4L^2}} \,. \label{eq:C4_res}
\ee
We recognize a  4-pt amplitude from $d$-dimensional \textit{flat space}. It is, up to a positive overall factor, the $S$-matrix element for the exchange of a  scalar with mass $m^2+\frac{(d-1)^2}{4L^2}$.  
In the next subsections we explain why these features happen. 

Finally we comment on the other singularities of ${\cal C}^{(4),\, {\rm ex}}_\pm (p_i)$. 
Singularities associated with the cubic subdiagrams occur if $\Delta_{1,2}$, or $\Delta_{3,4}$, and $\Delta_q$ simultaneously go to zero. The threshold in $\Delta_q$ requires $-q^2>0$ to be reached, which occurs in case of an $s$-channel. These subdiagram divergences correspond to  those of the  3-pt holographic vertices, Eq.\,\eqref{eq:C3}. 
They thus appear in both the 1PI term ${\cal C}^{(4),\, {\rm ex}}_{\pm} $ and the boundary term ${\cal C}^{(4),\, {\rm ex}}_{\pm, {\partial}} $.

\subsubsection{Singularities }

A pattern  emerges upon inspection of the perturbative boundary correlators. The singularities are associated with the scaling dimensions flowing through each vertices of the diagrams. In particular, the correlator always features a singularity at vanishing  total scaling dimension $\Delta_T\to 0$. 
Let us prove this last point for a general diagram. 

We consider the general structure  for the generic brane correlator given in Eq.\,\eqref{eq:C_gen}, that we reproduce here: 
\be
{\cal C}^{(l)}(p_i)= \left[\prod_{j=1}^l \int d\y_{j}  K(p_j,\y_j) \right] {\cal A}^{(l)}(p_{i},\y_{i}).
\label{eq:Cl}
\ee
Here for concreteness we focus on the LD$_-$ region and drop the $-$ subscript. The result in the LD$_+$ region is identical.

We  notice that the ${\cal A}^{(l)}$
subdiagram has similar structure as the bulk correlator described in \eqref{eq:M_LD_flat}, except that the external legs are amputated. More precisely, using the same steps as in Sec.~\ref{se:LD_correlator_gen} we find
\be
{\cal A}^{(l)}_{{\rm LD}_{d+1}}(p_i,\y_i)={\cal A}^{(l)}_{{\rm flat}_{d+1}}(p_i,\y_i) \prod_{i=1}^l\left(\frac{\y_i}{\L}\right)^{\frac{d-3}{2}}\,  \label{eq:A_LD_flat}
\ee
where ${\cal A}^{(l)}_{{\rm flat}_{d+1}}(\y_j)$ is an  amputated  correlator in flat space, 
 expressed in momentum-position coordinates $\left(p^\mu,\L\log(L/\y)\right)$.    It is invariant under the dilatation transformation. 
The scaling dimension of ${\cal A}^{(l)}_{{\rm LD}_{d+1}}$
is thus $-l\frac{d-3}{2}$, as determined by the powers in \eqref{eq:A_LD_flat}.\,\footnote{
An easy way  to  understand the scaling dimension of ${\cal A}^{(l)}_{{\rm LD}_{d+1}}$ is as follows. Start from  the bulk correlator \eqref{eq:M_LD_flat}, which has scaling dimension $l\frac{d-1}{2}$.  Each  amputation  removes  two powers of $\Phi$ and one integral in $r$. This amounts to a shift of the scaling dimension by $1-d$ and $1$ respectively. 
Doing this for the $l$ legs amounts to shifting the scaling dimension by $+l(2-d)$, which reproduces the scaling dimension of ${\cal A}^{(l)}_{{\rm LD}_{d+1}}$. 
}

We introduce the conformal coordinates, $z=L \log(L/\y)$ and Fourier transform ${\cal A}^{(l)}_{{\rm flat}_{d+1}}$
with respect to all $z_i$, 
{
\be
{\cal A}^{(l)}_{{\rm flat}_{d+1}}(p_j,\y_j) = \hat {\cal A}^{(l)}_{{\rm flat}_{d+1}}(p_j,z_j) = 
\int \frac{dp_{j}^z}{2\pi} e^{ip_j^z z_j}  \tilde {\cal A}^{(l)}_{{\rm flat}_{d+1}}(p_j,p^z_j)\,\delta\left(\sum p^z\right)\,. \label{eq:Aflat}
\ee
}
We have extracted the overall delta function which results from the translation invariance in  $z$ coordinates. 

We introduce Eq.\,\eqref{eq:Aflat} in the complete expression of the brane correlator ${\cal C}^{(l)}(p_i)$, Eq.~(\ref{eq:Cl}), and integrate over the $z_j$ coordinates. The outcome, and the subsequent steps leading to the final form, are 
{
\begin{align}
{\cal C}^{(l)}(p_j)  & =  {\cal N}_{\Delta_T} \left[ \prod_{i=1}^l \int \frac{dp_i^z}{2\pi} \frac{1}{ \Delta_i-i p^z_i} \right]   \tilde {\cal A}^{(l)}_{{\rm flat}_{d+1}}(p_j,p^z_j)\,\delta\left(\sum p^z\right)  \label{eq:C_step1} \\
& = {\cal N}_{\Delta_T}  \left[ \prod_{i=1}^{l-1} \int \frac{dp_i^z}{2\pi} \frac{1}{\Delta_i-i p^z_i} \right] \frac{1}{\Delta_l+i(\sum^{l-1}_{k=1} p_k)}  \tilde {\cal A}^{(l)}_{{\rm flat}_{d+1}}(p_j,p^z_j) \label{eq:C_step2} \\
& =  {\cal N}_{\Delta_T} \frac{1} {{\sum^l_{k=1} \Delta_k} } \tilde {\cal A}^{(l)}_{{\rm flat}_{d+1}}(p_j,-i\Delta_j) \,,  \label{eq:C_step3}
\end{align}
where we use the notation $\Delta_i\equiv \Delta_{p_i}$,}
with the overall factor
 \be
{\cal N}_{\Delta_T}=\frac{L^{l(2-\frac{d}{2})}}{r_b^{l(1-\frac{d}{2})}}
 \left(\frac{L}{r_b}\right)^{\Delta_T}\,.
 \ee

From   line \eqref{eq:C_step1} to \eqref{eq:C_step2}  we integrated the $\delta$ function in $p^z_l$. 
From   line \eqref{eq:C_step2} to \eqref{eq:C_step3} 
we performed the remaining integrals by closing the contour downward to pick the single poles, $p^z_i=- i \Delta_i$. The  pole outside the bracket lies upward in the complex $p_z$ plane and is thus not picked when closing the contours. 
We see that the combination $\sum^l_{k=1} \Delta_k\equiv \Delta_T$ appears in the denominator. Thus our calculation makes it clear that the presence of the $1/\Delta_T$ pole 
follows from  conservation of $p^z$. It is thus a consequence of the translation invariance in $z$ of $\hat {\cal A}^{(l)}_{{\rm flat}_{d+1}}(p_j,z_j)$, or equivalently of the invariance under dilatation of $ {\cal A}^{(l)}_{{\rm flat}_{d+1}}(p_j,r_j)$. 

Finally, taking the limit of simultaneous kinematic threshold $\Delta_i\to 0$, the residue of the $\Delta_T= 0$ pole is  
\be
\tilde {\cal A}^{(l)}_{{\rm flat}_{d+1}}(p_j,0)\big|_{p^2_j=m^2+\frac{(d-1)^2}{4L^2}}\,.
\ee
This residue is precisely the (on-shell) $S$-matrix element in $d$-dimensional space. 
This explains the residues obtained for the 3-pt and 4-pt diagrams \eqref{eq:C3}, \eqref{eq:C4_res}. The ${\cal N}_T$ factor reproduces the overall factors in the residues.\,\footnote{To compare the overall power of $L$ one needs to take into account that some factors of $L$ are included into $\tilde {\cal A}_{\rm flat}$.  }

Our analysis shows that the existence of the singularity and its peculiar residue are a consequence of the dilatation symmetry of the EFT.

    \subsection{Connection to the Cosmological Bootstrap}

The intriguing structure emerging from the brane correlators in the LD background is remini\-scent of the one appearing in  the wavefunction of the Universe in dS$_3$ space. 

Studies of the wavefunction of the Universe frequently use a flat space toy model with diagrams ending on a constant time hypersurface \cite{Arkani-Hamed:2017fdk,Arkani-Hamed:2018bjr, Benincasa:2018ssx, Baumann:2021fxj}. 
% The features of the wavefunctions in the flat space toy model are similar to those in  dS.
A singularity structure is observed in the wavefunction toy model, which is identical to the one obtained in our LD correlators  upon substituting  the scaling dimensions $\Delta_i$ with energies $E_i$. 
The fact that the residue at $\Delta_T=0$ is a flat space amplitude was also noticed, see {e.g.}~\cite{Arkani-Hamed:2017fdk}.  
The similarity between  the LD correlators and the wavefunction toy model comes from the fact that the spacetime backgrounds share similar symmetries.

To exhibit the connection between these seemingly very different backgrounds, let us first discuss the case of a brane in flat space. 
We have seen that the  dilatation symmetry constrains the form of the  brane correlators. Taking $L\to \infty$, LD$_{d+1}$ space becomes the half flat space with a boundary at $z=0$. In this limit the dilatation symmetry reduces to the one-dimensional translation symmetry in the $z$ coordinate (see Sec.~\ref{se:sym_prop}). This translation symmetry constrains the flat space correlators, just like  dilatation symmetry constrains the LD correlators. 
In the $L\to \infty$ limit,
 $\Delta_i/L$ tends to the  momentum transverse to the brane, $p^z_i$. 
Accordingly,  $\Delta_T/L$ of a given diagram   reduces to the sum of the external $p^z$ momenta.

The wavefunction toy model is  the spacelike version of a  brane in flat space.  The coordinate transverse to the hypersurface is time, and accordingly the conjugate variable is energy.  The correlators are analogous to those for a brane in flat spacetime upon substituting $p_i^z\to E_i$. This is why the singularities in this model involve  sums of energies, $\sum_i E_i$, instead of $ \sum_ip_i^z$ as for  a brane in flat spacetime, and $\sum_i\Delta_i$ for a  brane in LD spacetime.

The connection between the LD brane correlators and the wavefunction coefficients is especially interesting because techniques have been developed to study the latter via {e.g.}~analytical bootstrap \cite{Baumann:2020dch,Baumann:2021fxj,Benincasa:2019vqr} and polytope representations \cite{Arkani-Hamed:2017fdk,Arkani-Hamed:2018bjr,Benincasa:2019vqr}.  In particular, the singularity structure (analogous to the one of the LD correlators) essentially fixes the wavefunction coefficients, and tools are developed to further bootstrap the wavefunction of the Universe  \cite{Baumann:2021fxj}. 
It is thus natural to expect that the LD brane correlators can analogously be constrained using their singularities. 

An important difference between the wavefunction coefficients and our LD correlators is that, in the former, the singularities appear only upon analytical continuation, while the singularities that appear in our case are physical.\,\footnote{Paraphrasing \cite{Baumann:2021fxj}: ``Singularities can be useful and can be real''. }  Thus it could make sense to take these singularities as input and use them to bootstrap the  holographic dual theory.
These developments are left for future investigation.

\section{Holography at Finite Temperature}
\label{se:finiteT}

We study the gravitational and thermodynamic behavior of the $d$-dimensional holographic theory in the presence of a planar bulk black hole.

\subsection{The Linear Dilaton Black Hole}

The general scalar-gravity action in $D$ dimensions in the presence of a brane {at $r=r_b$} is defined in section \ref{se:correlators}, see Fig.\,\ref{fig:LD_brane}. Here we solve the  field equations   allowing for the existence of a black hole in the metric,
\be
ds^2=g_{MN}dx^Mdx^N= e^{-2A(\y)} \left( -f(r) d\tau^2 + d\x_{D-2}^2 \right) + \frac{e^{-2B(\y)}}{f(\y)} d\y^2\,,\label{eq:ds2BH}
\ee
where $f(r)$ is the blackening factor. The metric describes an event horizon at {$r=r_h$ if $f(r_h)=0$}, and if the $A,B$ metric factors and the dilaton are regular on this surface. The temperature associated with the horizon is {$T_h = e^{B(r_h) - A(r_h)} |f^\prime(r_h)| / 4\pi $}, see Eq.~(\ref{eq:Th_general}).

\begin{figure}[t]
\centering
\includegraphics[trim={0cm 3cm 6cm 7cm},clip,width=0.7\textwidth]{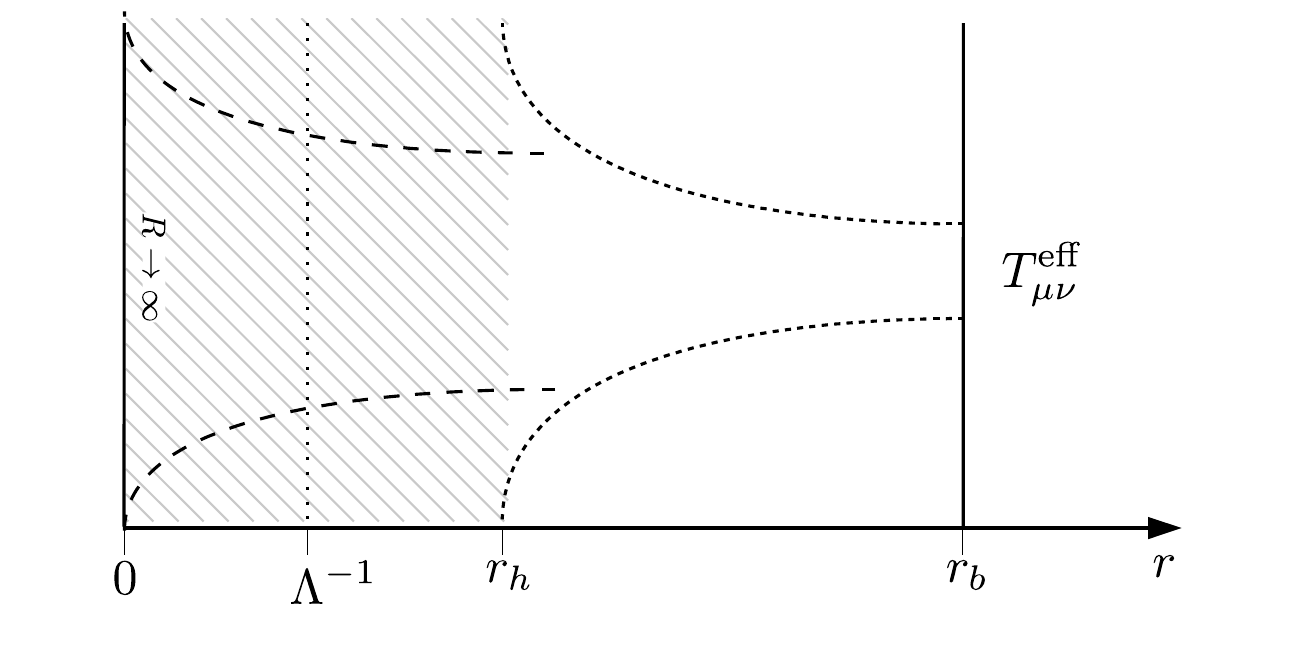}
\caption{\it  The linear dilaton spacetime with a brane and a bulk black hole. The projection of the black hole horizon onto the brane gives rise to an effective stress tensor $T^{\rm eff}_{\mu\nu}$. }
\label{fig:LD_finiteT}
\end{figure}

The independent field equations for the $D$-dimensional background are~\cite{Megias:2018sxv,Fichet:2022ixi,Fichet:2022xol} 
\begin{eqnarray}
\hspace{-0.3cm}&& \frac{f^{\prime\prime}(r)}{f^\prime(r)} + B^\prime(r) - (D-1) A^\prime(r) = 0 \,,  \label{eq:EoMT1}  \\
\hspace{-0.3cm}&& A^{\prime\prime}(r) + A^\prime(r) B^\prime(r) - \bar \phi^\prime(r)^2 = 0 \,,  \label{eq:EoMT2}  \\
\hspace{-0.3cm}&&A^\prime(r)^2 + \frac{1}{D-1} \left(  2 \frac{e^{-2 B(r)}}{f(r)} \bar 
V(\bar\phi) - \frac{f^\prime(r)}{f(r)} A^\prime(r) - \bar\phi^\prime(r)^2  \right) = 0 \,.    \label{eq:EoMT3}
\end{eqnarray}
There is a fourth field equation  which is redundant, as it can be expressed in terms of these three equations (see App.~\ref{sec:app_T} for a discussion). This system admits the following solution
\begin{equation}
ds^2 = \frac{1}{\eta^2 r_b^2}\frac{dr^2}{f(r)} + \frac{r^2}{L^2} \left( - f(r) d\tau^2 + d\x_{D-2}^2 \right) \,, \label{eq:ds2BH2}
\end{equation}
\begin{eqnarray}
&&f(r) = 1 - \left( \frac{r_h}{r}\right)^{D-2} \,,  \label{eq:fr}\\
&&\bar \phi(r) = \log \left( \frac{\y_b}{\y} \right) + \bar v_b \,, \label{eq:phirb}
\end{eqnarray}
where the scale $\eta$ is defined by
\begin{equation}
\eta \equiv k \, e^{\bar v_b} \,. \label{eq:eta}
\end{equation}

As in the zero temperature case, the $c$ parameter is fixed to $c = \frac{1}{\eta r_b}$
because $\frac{\partial \bar v_b}{\partial r_b}=0$ (see section \ref{se:halves}). This constraint  implies for instance the fundamental property that the $d$-dimensional Planck scale is independent of $r_b$,  ensuring the consistency of gravity in the $d$-dimensional theory.  See Sec.~\ref{se:conditions} for other implications.

The curvature singularity at $r=0$ gets censored by the black hole horizon at $r_h>0$, as expected from a good singularity. Moreover, assuming the EFT of gravity breaks down at distances $\Delta x \sim \frac{1}{\Lambda}$, we require \be r_h\gtrsim \frac{1}{\Lambda} \,.\ee
In turn, the horizon censors the region where the EFT of gravity would break down. 
The geometry is summarized in Fig.~\ref{fig:LD_finiteT}.

The fundamental domain of the $r$ coordinate is $r\in [0,r_b]$, so that, in the EFT $r\in [\Lambda^{-1}, r_b ]$, as shown in Fig.\,\ref{fig:LD_finiteT}. For convenience, throughout this section we use a $\mathbb{Z}_2$ orbifold convention as in \cite{Shiromizu:1999wj}
which implies that the spacetime is mirrored on the other side of the brane. The domain of integration of $r$ is thus doubled in the action. A notable implication is that the entropy of the black hole horizon is doubled when using this convention, see {e.g.}~Ref.~\cite{Megias:2018sxv}.

\paragraph{Black hole thermodynamics.}
{Here we summarize some results from the thermodynamic of the black hole. For more details about the calculation, see App.~\ref{sec:app_T}.}

The Hawking  temperature of the black horizon is
\begin{equation}
T_h = \frac{D-2}{4\pi}    \frac{r_b}{L} \eta\,. \label{eq:Th}
\end{equation}
The entropy  of the black hole may be computed by using a Bekenstein-Hawking-type entropy formula~\footnote{{Where we have introduced an extra factor of 2 from the $\mathbb Z_2$ orbifold.}}
\begin{equation}
S_h  = \frac{\mathcal A}{ 2 G_D} \label{eq:Sh}  \,,
\end{equation}
{where $G_D \equiv \frac{M_D^{2-D}}{8\pi}$ is the $D$-dimensional Newton constant.}
We introduce the entropy density per unit of comoving volume, $s_h \equiv \frac{S_h}{V_{D-2}}$  with  $V_{D-2} = \int d^{D-2}x$. 
Using the metric solution Eq.~(\ref{eq:ds2BH2}), we find 
\begin{equation}
s_h  = \frac{1}{ 2 G_D} \left(  \frac{r_h}{L} \right)^{D-2} \,.
\label{eq:sh}
\end{equation}

 Notice that the entropy density depends only on the horizon position $r_h$ while the temperature $T_h$ is a function of the brane position $r_b$. Therefore the black hole entropy and  temperature are independent of each other.   This is a special property of the LD background, which  does not occur in general in other scalar-gravity systems (see  \cite{Gursoy:2008za} for related results in $D=5$).

\subsection{The Effective Einstein Equations}
\label{subsec:DM_LD}

We now turn to the gravitational behavior of the theory projected onto the brane.
The induced metric on the brane {at $r=r_b$} is
\begin{equation}
d\bar s^2 = \bar g_{\mu\nu} dx^\mu dx^\nu = -dt^2 + e^{-2A(r_b)} d\x_{D-2}^2 \,,  \label{eq:metric_brane}
\end{equation}
where the brane proper time is $dt = e^{-A(r_b)} \sqrt{f(r_b)} \, d\tau$. Notice the $e^{-A(r_b)}$ amounts to a spatial scale factor. The metric \eqref{eq:metric_brane} can be understood as a generalization of the FRW metric to arbitrary dimension.

To study gravity in the holographic theory, we compute  the effective $d$-dimensional Einstein equations as seen by a brane observer. 
These are computed from the $D$-dimensional Einstein equations by projecting on the brane via the Gauss equation and the Israel junction condition (see \cite{Shiromizu:1999wj} for the original calculation in AdS$_5$). We recall we use the same orbifold convention as in~\cite{Shiromizu:1999wj}.

We find that the effective Einstein equations have the form
\begin{equation}
{}^{(d)}G_{\mu\nu} = \frac{1}{M_{d}^{d-2}} \left( T_{\mu\nu}^b + T_{\mu\nu}^{\rm eff}\right) + O\left( \frac{T_b^2}{M_D^{2D-4}} \right) \,, \label{eq:D_1_Einstein}
\end{equation}
where $T_{\mu\nu}^b$ is the stress tensor of possible brane-localized  matter. 
The indices in \eqref{eq:D_1_Einstein} are contracted with the induced metric \eqref{eq:metric_brane}. Equation \eqref{eq:D_1_Einstein} has the form of the standard Einstein equations with an {extra} effective stress tensor $T^{\rm eff}_{\mu\nu}$. 
Moreover, it turns out that the structure of the $T^{\rm eff}_{\mu\nu}$ tensor represents a 
$d$-dimensional \textit{perfect fluid} at rest:\,\footnote{
 The stress-energy tensor of a {perfect fluid} is in general expressed as~\cite{Kovtun:2012rj}
\begin{equation}
T_{\mu\nu} = (\rho + P) U_\mu U_\nu + P g_{\mu\nu} \,,
\end{equation}
where $\rho$ is the energy density, $P$ is the pressure, while $U^\mu = \gamma (1, v^i)$ is the local fluid velocity with $\gamma = 1/\sqrt{1 - \vec{v}^2}$ the Lorentz factor. The perfect fluid at rest, {i.e.}~in a comoving reference frame, has
$U^\mu = (1,\vec{0})$.
The fact that we obtain a fluid at rest is tied to our choice of a static black hole ansatz in the metric. 
Boosted black hole solutions are also considered in the context of the 
fluid/gravity correspondence \cite{Bhattacharyya:2007vjd} in order to compute higher order corrections to the constitutive relations of the fluid, which involve viscosities and other transport coefficients. }
\begin{equation}
T^{\rm eff,\mu}_\nu = g^{\mu\lambda} T^{\rm eff}_{\lambda \nu} =  \textrm{diag}(-\rho, P, \cdots, P) \,.
\end{equation}

In our notation, the  effective $d$-dimensional cosmological constant is included in $T^{\rm eff}_{\mu \nu}$.
We  work in the low energy regime
\begin{equation}
|T^b_{\mu\nu}| \ll \frac{M_D^{2D-4}}{M_d^{d-2}} \,, \label{eq:LE_regime}
\end{equation}
such that the higher-order terms in Eq.~(\ref{eq:D_1_Einstein}) can be neglected. This implies $\rho_b \ll M_{d}^{d-2} \eta^2$, so that $\rho_b$ is negligible as compared to $\rho_{\rm eff}$.

\subsubsection{Holographic fluid from  warped spacetime}

\label{se:fluid_warped}

We present the result for the effective stress tensor obtained from the general metric \eqref{eq:ds2BH}.  The  effective stress tensor contains three  contributions
\begin{equation}
T_{\mu\nu}^{\rm eff} = \tau^W_{\mu\nu} + \tau^\phi_{\mu\nu} + \tau^\Lambda_{\mu\nu} \,,
\end{equation}
which are as follows.
\begin{itemize}
\item[{\it (i)}] The $\tau^W_{\mu\nu}$ is from the projection of the $D$-dimensional Weyl tensor ${}^{(D)}C_{MNPQ}$ onto the brane, 
\begin{equation}
\frac{1}{M_{d}^{d-2}} \tau^W_{\mu\nu} = -{}^{(D)}C^M{}_{NPQ} \, n_M n^P \bar g_\mu{}^N \bar g_\nu{}^Q \,,
\end{equation}
where $n^M$ is the unit vector normal to the brane, {i.e.}
\begin{equation}
n^M = \left( 0,\cdots,0,\sqrt{f(r)} e^{B(r)}  \right) \,, 
\end{equation}
and $\bar g_{MN} = g_{MN} - n_M n_N$ is the induced metric on the brane.  This leads to the following values for the energy density $\rho^W$ and pressure $P^W$:
\begin{equation}
\rho^W(r_b) = (d-1) P^W(r_b) =  - \frac{(d-1)(d-2)}{2d} M_{d}^{d-2}  \,  e^{2 B(r_b)} A^\prime(r_b) f^\prime(r_b)  \,.
\end{equation}
\item[{\it (ii)}] The  {$\tau^\phi_{\mu\nu}$} is from the projection of the bulk stress tensor,
\begin{align}
\frac{1}{M_{d}^{d-2}} \tau^\phi_{\mu\nu} &= \frac{(d-2)}{(d-1) } \frac{1}{M_D^{D-2}} \left[ T^\phi_{MN} \bar g_\mu{}^M \bar g_\nu{}^N + \left( T^\phi_{MN} n^M n^N  - \frac{1}{d} T^{\phi,\, M}_M  \right) \bar g_{\mu\nu} \right] \,, \\
&= \frac{(d-1)(d-2)}{d} \left(  \frac{1}{2} e^{2B(r_b)} f(r_b) \bar\phi^\prime(r_b)^2 - \bar V \right) \bar g_{\mu\nu} \,.
\end{align}
This leads to the following values of $\rho^\phi$ and $P^\phi$:
\begin{equation}
\rho^\phi(r_b) = - P^\phi(r_b) =  - \frac{(d-1)(d-2)}{2d} M_d^{d-2}  e^{2B(r_b)} A^\prime(r_b) \left[ d \, f(r_b) A^\prime(r_b) - f^\prime(r_b) \right]     \,.
\end{equation}
\item[{\it (iii)}] The  $\tau^\Lambda_{\mu\nu}$ is the contribution from the brane tension,
\begin{equation}
\frac{1}{M_{d}^{d-2}} \tau^\Lambda_{\mu\nu} = - \frac{(d-2)}{8 (d-1)} \frac{\Lambda_b^2}{M_D^{2(D-2)}} \bar g_{\mu\nu}  \,,
\end{equation}
which yields the values
\begin{equation}
\rho^\Lambda = - P^\Lambda = \frac{(d-2)}{8 (d-1)} \frac{M_{d}^{d-2}}{M_D^{2(D-2)}}  \Lambda_b^2 \,,
\end{equation}
where $\Lambda_b$ is the brane tension with mass dimension equal to $d$, defined in the brane action Eq.\,\eqref{eq:action_brane}.
\end{itemize}

\subsubsection{Holographic fluid from  LD spacetime}

\label{se:fluid_LD}

We now plug the linear dilaton black hole solutions of Eqs.~(\ref{eq:ds2BH2})-(\ref{eq:phirb}) into the general expressions of section \ref{se:fluid_warped}.   We find the effective energy density 
\begin{equation}
\rho_{\rm eff}(r_b) = \rho^W(r_b) + \rho^\phi(r_b) + \rho^\Lambda \,,
\label{eq:rhoeff_sum}
\end{equation}
with
\begin{align}
\rho^W(r_b) &=  \frac{(d-1)^2 (d-2)}{2d} M_{d}^{d-2} \, \eta^2 \left( \frac{r_h}{r_b}\right)^{d-1} \,, \\
\rho^\phi(r_b)  &=  \frac{(d-1)(d-2)}{2d} M_{d}^{d-2} \, \eta^2 \left(  \frac{r_h}{r_b}\right)^{d-1} - \frac{1}{2}(d-1)(d-2) M_{d}^{d-2} \, \eta^2 \,, \\
\rho^\Lambda &=  \frac{1}{2}(d-1)(d-2) M_{d}^{d-2} \, \eta^2   + M_{d}^{d-2} \Lambda_{d}  \,,
\end{align}
where we have introduced the $d$-dimensional cosmological constant $\Lambda_d$  {(with mass dimension equal to $2$)} and the Planck scale $M_d$. These are identified in the effective Einstein equations  as 
\begin{equation}
\Lambda_{d} =   - \frac{(d-1)(d-2)}{2}  \eta^2 + \frac{d-2}{8 (d-1)} \frac{\Lambda_b^2}{M_D^{2D-4}}  \,, \label{eq:Lambda_LD}
\end{equation}
\begin{equation}
\frac{M_D^{D-2}}{M_{d}^{d-2}} = \frac{1}{2} (d-2) \eta \sqrt{ 1 + \frac{2}{(d-1)(d-2)} \frac{\Lambda_{d}}{\eta^2} } \,. \label{eq:MD_LD}
\end{equation}

It turns out  that the first term of $\rho^\Lambda$ cancels the constant contribution in $\rho^\phi(r_b)$ in the sum \eqref{eq:rhoeff_sum}, so that the effective energy density is given by
\begin{equation}
\rho_{\rm eff}(r_b) = \frac{1}{2} (d-1)(d-2) M_{d}^{d-2} \, \eta^2 \left( \frac{r_h}{r_b}\right)^{d-1} + M_{d}^{d-2} \Lambda_{d}  \,. \label{eq:rho_eff} 
\end{equation}
On the other hand, we find the following results for the pressure,
\begin{align}
P^W(r_b) + P^\phi(r_b) &= \frac{1}{2} (d-1)(d-2) M_{d}^{d-2} \, \eta^2 \,, \\
P^\Lambda &= -\frac{1}{2} (d-1)(d-2) M_{d}^{d-2} \, \eta^2  - M_{d}^{d-2} \Lambda_{d} \,,
\end{align}
so that the summation of all the contributions yields \be P_{\rm eff} = - M_{d}^{d-2} \Lambda_{d}\,. \ee

This result can be interpreted as a {\it pressureless} perfect fluid living in a $d$-dimensional spacetime with a non-vanishing vacuum energy density $\Lambda_d$, {i.e.}
\begin{equation}
\rho_{\rm eff}(r_b) = \rho_{\rm fluid}(r_b) + \rho_{\rm vacuum} \,, \qquad P_{\rm eff}(r_b) = P_{\rm fluid}(r_b) + P_{\rm vacuum} \,, \label{eq:rho_fv}
\end{equation}
with
\begin{align}
\rho_{\rm fluid}(r_b) &= \frac{1}{2} (d-1)(d-2) M_{d}^{d-2} \, \eta^2 \left( \frac{r_h}{r_b}\right)^{d-1} \,, \quad P_{\rm fluid}(r_b) = 0 \,, \label{eq:rho_P_fluid}\\
\rho_{\rm vacuum} &= -P_{\rm vacuum} = M_d^{d-2} \Lambda_d \,. \label{eq:rho_P_vacuum}
\end{align}

In summary, the LD geometry induces  a  holographic theory which  contains a pressureless fluid.  This is in contrast with AdS$_{d+1}$  geometry, where  carrying out an analogous  calculation leads to a holographic fluid with nonzero pressure {$P^{\rm AdS}_{\rm fluid}=\frac{1}{d-1}\rho^{\rm AdS}_{\rm fluid}$,} in accordance with $d$-dimensional conformal invariance.\,\footnote{
{In a brane cosmology application, the pressureless fluid from LD$_5$ could play the role of dark matter~\cite{Fichet:2022ixi,Fichet:2022xol}, while the holographic fluid from AdS$_5$  would instead behave as  dark radiation.} }

In the following subsections we  study the consistency of these results, as well as some of their physical consequences.

\subsection{Thermodynamics of the Holographic Fluid }
\label{subsec:brane_thermodynamics}

A thermodynamic description of the $d$-dimensional theory can be inferred from the effective energy density derived in Eq.~(\ref{eq:rho_eff}). {  The physical parameters of the system  are  the horizon location $r_h$ and the brane location $r_b$. 
We introduce thermodynamical variables which are understood as functions of these parameters. 
} 

First we define the volume and temperature of the system. The spatial volume of the brane is given  by
\begin{equation}
V_b = V_{d-1} \left(  \frac{r_b}{L}\right)^{d-1}  \,, \label{eq:Vb}
\end{equation}
where $V_{d-1} = \int d^{d-1}x $ is the brane comoving volume.  
The  black hole  temperature on the brane is  
\begin{equation}
T_b = T_h\frac{L}{r_b} = \frac{(d-1)}{4\pi} \eta \,, \label{eq:Tb}
\end{equation}
where  $T_h$ is the horizon temperature (see Eq.~(\ref{eq:Th_general})). 
The temperature on the brane is obtained from the horizon temperature by multiplying  it by $1/\sqrt{|g_{\tau\tau}|}$~\cite{Hawking:1982dh}. 
It turns out that $T_b$ is independent of $r_b$, and thus from $V_b$.~\footnote{{In the present analysis we are assuming $r_h \ll r_b$, so that the $r_b$ dependence of $T_b$ induced by the $1/\sqrt{f(r_b)}$  factor is  a small correction that we  neglect.}} {Moreover $T_b$ is also independent of $r_h$, it is thus a universal constant of the $d$-dimensional theory.  This is a special property of the LD background. {We will see later that $T_b$ can be identified with the Hagedorn temperature.}}

By the fundamental laws of thermodynamics, the reversible variation of total energy $E$ of the system should satisfy 
\begin{equation}
dE = T dS-PdV \,. \label{eq:law_thermodynamics}
\end{equation}
Using that $E$, $S$ and $V$ are $0$-forms and $d( d X) = 0$ for any $k$-form $X$, taking the exterior derivative of \eqref{eq:law_thermodynamics} gives $dT \wedge dS - dP \wedge dV = 0$, where $\wedge$ denotes the exterior product. Considering that: \textit{i)} $T_b$ is independent on $r_h$ and $r_b$ which implies $d T_b = 0$, and \textit{ii)} $V_b$ depends only on $r_b$; it follows  that $P_{\rm fluid}$ depends at most on $r_b$. Finally $P_{\rm fluid}(r_b)$ should vanish when $r_h \to 0$ since the black hole ceases to exist in that limit, hence $P_{\rm fluid} = 0$. We thus recover from thermodynamics the vanishing pressure that we found in Sec.~\ref{se:fluid_LD}. This provides a consistency check of our thermodynamic approach.

{On the other hand,} using the energy density $\rho_{\rm fluid }$ found in \eqref{eq:rho_P_fluid}, it turns out that $E_{\rm fluid} = \rho_{\rm fluid } V_b $ is independent of $r_b$. 
Therefore, we get from the relation~(\ref{eq:law_thermodynamics})
\be
\left[ \frac{\partial E_{\rm fluid }}{\partial r_h} -  T_b \frac{\partial  S_{\rm fluid }}{\partial r_h} \right] d r_h  + \left[ P_{\rm fluid } \frac{\partial V_b}{\partial r_b}  - T_b \frac{\partial S_{\rm fluid }}{\partial r_b} \right]d r_b  = 0 \label{eq:dE_fluid}
\,,
\ee
where we have kept for the moment the contribution of the pressure. Notice that
 each bracket must separately vanish. From the second bracket in Eq.~(\ref{eq:dE_fluid}) we see that the condition $P_{\rm fluid}=0$ implies that $ S_{\rm fluid}$ is independent of $r_b$.  
Moreover, from the first bracket in Eq.~(\ref{eq:dE_fluid}) we {similarly} infer that $E_{\rm fluid}$ and $S_{\rm fluid}$ are proportional to each other, up to an additive integration constant.  We set the  additive  constant to zero using that $S_{\rm fluid}$ and $E_{\rm fluid}$ should both vanish in the $r_h\to 0$ limit, for which the black hole does not exist.

In sum, we obtain from \eqref{eq:dE_fluid} the relation
\be E_{\rm fluid} = T_b \, S_{\rm fluid}\,. \label{eq:SpropE} \ee
   This automatically implies that  the fluid free energy {$F=E-TS$} vanishes, $F_{\rm fluid}=0$.~\footnote{An alternative proof of Eq.~(\ref{eq:SpropE}) can be done by using $dF = -S dT - PdV$. Taking the exterior derivative of this relation and using similar arguments, one concludes that $P_{\rm fluid} = 0$ and then $d F_{\rm fluid} = 0$. Hence $F_{\rm fluid}$ is a constant in $r_h$ and $r_b$ that should vanish, and so $E_{\rm fluid} = T_b \, S_{\rm fluid}$. This proof uses only that $T_b$ is independent on $r_h$ and $r_b$, and $V_b = V_b(r_b)$.}  Finally, the total entropy is deduced from Eq.\,\eqref{eq:SpropE}  by using the value of the temperature in \eqref{eq:Tb}. The total energy and entropy are 
\begin{eqnarray}
E_{\rm fluid} &=&   \frac{1}{2} (d-1) (d-2) \eta^2 M_{d}^{d-2} \left(  \frac{r_h}{L} \right)^{d-1} V_{d-1}  \,, \\
S_{\rm fluid} &=&   2 \pi(d-2) \eta M_{d}^{d-2} \left( \frac{r_h}{L} \right)^{d-1}  V_{d-1}  \,. \label{eq:Sfluid}
\end{eqnarray}

 We notice that $S_{\rm fluid}$ can be independently derived  from  the Bekenstein-Hawking entropy of the black hole,  Eq.\,\eqref{eq:Sh}.  Starting from 
Eq.\,\eqref{eq:sh}, we multiply by the redshift factor to get the entropy density at the brane, {i.e.}~$s_b = s_h \left( L/r_b \right)^{d-1}$. We use  the relation between $M_D^{D-2}$ and $M_d^{d-2}$ given by Eq.~(\ref{eq:MD_LD})  assuming  $\Lambda_d \ll \eta^2$, so that the contribution from $\Lambda_d$ is negligible.  The result precisely reproduces Eq.\,\eqref{eq:Sfluid}:
\begin{equation}
S_b  = s_b V_b =  4\pi M_D^{D-2} \left( \frac{r_h}{L} \right)^{d-1} V_{d-1} = S_{\rm fluid}\,,
\end{equation}
where we have used the expression of $V_b$ given by Eq.~(\ref{eq:Vb}). 
From a comparison with Eqs.~(\ref{eq:Sh})-(\ref{eq:sh}) one easily realizes that $S_b = S_h$, which  is a consequence of the independence of $S_b$ on $r_b$. Hence the entropy of the holographic fluid matches exactly the black hole entropy, $S_{\rm fluid}=S_{h}$.

Summarizing, we have found that the holographic fluid  has a universal temperature, a vanishing pressure and has 
$S_{\rm fluid}\propto E_{\rm fluid}$. 
These peculiar features
{reproduce} the   \textit{Hagedorn  behavior} \cite{Hagedorn:1965st} that 
typically appears in string thermodynamics \cite{Salomonson:1985eq,Atick:1988si} and is, in particular, obtained for LST~\cite{Maldacena:1996ya,Maldacena:1997cg,Harmark:2000qm,Kutasov:2000jp,Parnachev:2005hh,Goykhman:2013oja}.\,\footnote{Identifying the brane temperature $T_b$, Eq.~(\ref{eq:Tb}), with the Hagedorn temperature $T_H$ from LST~\cite{Harmark:2000qm} given by $T_H=(2\pi \sqrt{\alpha' N})^{-1}$,  we obtain $\eta = 5/\sqrt{4\alpha' N}$.   $\alpha'$ is the string tension, $N$ is the number of NS5-branes, and  we have set $d=6$. } This  behavior  also occurs for CFT$_2$ with $T\overline T$ deformations \cite{Giveon:2017nie}.
Here the same phenomenon arises in our low-energy approach, from 
the projection of the bulk physics onto the brane.   The Hagedorn behavior appears generically for any $d>2$.

\subsection{Time Evolution of the Holographic Fluid}
\label{subsec:time_evolution}

Further nontrivial checks of  the fluid properties obtained in
Secs.~\ref{se:fluid_LD} and \ref{subsec:brane_thermodynamics}
can be achieved by 
studying the time evolution of the brane. 
This is, in a sense, a ``cosmological'' study of the $d$-dimensional linear dilaton  braneworld. 

We allow  the brane to evolve with time in the bulk: $r_b=r_b(t)$. 
The effective Einstein equations Eq.~(\ref{eq:D_1_Einstein}) give the   first $d$-dimensional Friedmann equation on the brane:
\begin{equation}
 \frac{1}{2} (d-1)(d-2) M_{d}^{d-2} H^2 = \rho_b + \rho_{\rm eff} + O\left( \frac{\rho_b^2}{\eta^2 M_{d}^{d-2}} \right) \,,  \label{eq:Friedmann_Eq}
\end{equation}
where {$H \equiv \dot a_b(t)/a_b(t)$ is the Hubble parameter with $a_b = e^{-A(r_b)}$}, and $\rho_b$ is the energy density of brane localized matter which is neglected as $\rho_b \ll \rho_{\rm eff}$. {When using the black hole solutions of
Eqs.~(\ref{eq:ds2BH2})-(\ref{eq:phirb}), one has $a_b(t) = r_b(t)/L$.} In the following the overdot means differentiation with respect to the brane proper time~$t$.

\paragraph{Time evolution}
Let us study the consistency of the pressureless fluid result of Sec.~\ref{subsec:DM_LD} from a comparison with the solution of the Friedmann equation. Given a perfect fluid in $d$ spacetime dimensions with equation of state $P = w \rho$, the solution of the Friedmann equation is
\begin{equation}
a_b(t) \propto t^{\frac{2}{(1+w)(d-1)}}  \qquad \textrm{with} \qquad  \rho \propto \frac{1}{a_b^{(1+w) (d-1)} } \,.
\end{equation}
It turns out that: \textit{i)} a radiation dominated universe $(w = \frac{1}{d-1})$ behaves as $a_b(t) \propto t^{\frac{2}{d}}$ with $\rho \propto 1/a_b^d$, \textit{ii)} a matter dominated universe $(w = 0)$  behaves as $a_b(t) \propto t^{\frac{2}{d-1}}$ with $\rho \propto 1/a_b^{d-1}$, and \textit{iii)} a universe dominated by the cosmological constant $(w = -1)$ behaves as $\log a_b(t) \propto t$ with $\rho \propto$ cte.  A comparison with Eq.~(\ref{eq:rho_P_fluid}) confirms that the  energy density $\rho_{\rm fluid}$ behaves as a pressureless  matter term in the $d$-dimensional Friedmann equation.\,\footnote{
For completeness, we provide the solution of the Friedmann equation including both the perfect fluid contribution and the cosmological constant term:
\begin{equation}
a_b(t) = a_h \left[  (d-1) \frac{\eta}{2\gamma} \sinh(\gamma(t-\lambda)) \right]^{\frac{2}{d-1}}  \,, \qquad \gamma = \sqrt{ \frac{(d-1)}{2(d-2)} \frac{\rho_{\rm vacuum}}{M_{d}^{d-2}} } \,, \label{eq:rbt}
\end{equation} 
{where $a_h \equiv e^{-A(r_h)}$} and $\lambda$ is an integration constant that can be freely chosen.  This formula reproduces  the power-law behavior for a matter dominated universe $(\rho_{\rm vacuum} \ll \rho_{\rm fluid}(r_b))$, and the exponential behavior for a universe dominated by the cosmological constant $(\rho_{\rm vacuum} \gg \rho_{\rm fluid}(r_b))$. }

\paragraph{Conservation equation}

The conservation equation of the $D$-dimensional bulk stress tensor ({i.e.}~the Bianchi identity) evaluated on the brane can be written as \cite{Tanaka:2003eg,Langlois:2003zb} 
\begin{equation}
\dot \rho_{\rm eff} + d H \rho_{\rm eff} + H T_\mu^{{\rm eff}\, \mu} =  - 2 \left( 1 + \frac{\rho_b}{\Lambda_b}\right) T^\phi_{MN} u^M n^N \,, \label{eq:conservation}
\end{equation}
where $u^M = (\dot \tau, \mathbf 0, \dot r_b)$ {with $\dot \tau = e^A \sqrt{f + \dot r_b^2 e^{-2B}}/f$} is the timelike unit vector for brane observers. The right-hand side of this equation represents the energy flux from the brane into the bulk. To get this expression we have used 
\begin{equation}
(d-2) \frac{M_{d}^{d-2}}{M_D^{D-2}} H T^\phi_{MN} n^M n^N + H \tau_\mu^{\Lambda  \, \mu} = H T_\mu^{{\rm eff} \, \mu} \,.
\end{equation}

 Using the explicit expression of $T^\phi_{MN}$ and after some non-trivial cancellations, it turns out that $T^\phi_{MN} u^M n^M = 0$ in the low-energy regime.  Finally, by using that
\begin{equation}
T_\mu^{{\rm eff}\, \mu}  = - d \cdot (\rho^\phi(r_b) + \rho^\Lambda )  = -\rho_{\rm fluid}(r_b) - d \cdot \rho_{\rm vacuum}\,,
\end{equation}
it is easy to verify that the conservation equation is satisfied by the effective energy density $\rho_{\rm eff}$ of Eqs.~(\ref{eq:rho_fv})-(\ref{eq:rho_P_vacuum}). A similar analysis to that performed above was done in Refs.~\cite{Fichet:2022ixi,Fichet:2022xol} for $d = 4$ and $\rho_{\rm vacuum} = 0$. The present results agree and generalize  those of the latter references.

Finally, the conservation of entropy is automatically ensured since ${\cal S}_{\rm fluid}$ is independent of $r_b$, and thus from the brane proper time.

\subsection{Remarks on the General Solution} 

\label{se:conditions}

We discuss in more details the fixing of the
parameter $c$ appearing in the general solution of the field equations {with no black hole~(\ref{eq:Ar_Br})-(\ref{eq:phir}), and in the presence of a planar black hole}~(\ref{eq:AT})-(\ref{eq:phibT}). 
We find that $c$ is fixed by various equivalent conditions. 

We first impose the bulk conservation equation Eq.\,\eqref{eq:conservation} ({i.e.}~the bulk Bianchi identity) and work in the low-energy regime as defined by Eq.\,\eqref{eq:LE_regime}. Then the following conditions are equivalent to each other:
\begin{enumerate}[label={\it \roman*)}]
    \item The dilaton vev on the brane {$\bar v_b \equiv \bar\phi(r_b)$} is independent of $r_b$, and thus does not vary in time; $\frac{d \bar v_b}{d r_b}=0$. 
    \item  Likewise for  the effective $d$-dimensional cosmological constant $\Lambda_d$; $\frac{d \Lambda_d}{d r_b}=0$. 
    \item  Likewise for the effective $d$-dimensional Planck scale {$M_d$}; $\frac{d M_d}{d r_b}=0$. 
    \item Likewise for  the effective pressure; $\frac{dP_{\rm eff}}{d r_b} = 0$. 
    \item Effective conservation equation; $\dot \rho_{\rm eff} + d H \rho_{\rm eff} + H T_\mu^{{\rm eff}\, \mu} = 0$. 
    \item   The energy loss from the brane into the bulk is negligible; $T^\phi_{MN} u^M n^N=\mathcal O(H^3)$. 
    \item $\C \propto \frac{1}{r_b^a}$. 
\end{enumerate}
 We can unambiguously fix the parameter $c$ by using the general solution for $\bar\phi(r)$, together with the definitions $\bar v_b \equiv \bar\phi(r_b)$ and $\eta \equiv k \, e^{\bar v_b}$. The result is
\begin{equation}
\C = \frac{|a|}{\eta L} \left(\frac{ L}{r_b} \right)^a  \,,
\end{equation}
and $\eta$ is independent of $r_b$ as is clear from  condition {\it i)}.

Notice that most of the above conditions are physically meaningful. Their equivalence is a compelling verification of the self-consistency of our results.

The $d$-dimensional Friedmann equation does not impose any additional constraint on the parameters of the solution. Using the general solutions {in the presence of a bulk black hole} (\ref{eq:AT})-(\ref{eq:phibT}), 
the solution of the Friedmann equation is given by Eq.~(\ref{eq:rbt}) where now $a_b(t) = (r_b(t) / L)^{a}$.  For $\Lambda_d = 0$ it is
\begin{equation}
a_b(t) = a_h \, \left[ (d-1) \frac{\eta}{2} (t-\lambda)\right]^{\frac{2}{d-1}} \,. \label{eq:ab_t}
\end{equation}
The power corresponds to $w = 0$, in agreement with the pressureless fluid result.

As a final remark, we mention that  the parameter $a$ remains unconstrained.  It can thus be freely chosen and the geometry does not depend on it, in accordance with the discussion in App.\,\ref{app:solutions}. This property becomes explicit, for instance, in the solution of the Friedmann equation provided in Eqs.~(\ref{eq:rbt}) and~(\ref{eq:ab_t}), which are independent of $a$.

\section{Summary and Outlook}

\label{se:con}

In this work we have studied the $(d+1)$-dimensional linear dilaton (LD) spacetime with a focus on holography. Let us recapitulate our results. 

The LD spacetime has a (good) curvature singularity but no boundary. Its conformal structure is the same as  Minkowski space. 
The timelike geodesics are attracted towards the singularity while 
the null geodesics behave very much as in flat space. 

The LD spacetime has symmetries besides the $d$-dimensional Poincaré invariance of the flat slicing: a conformal dilatation symmetry and a conformal inversion symmetry acting on the warped coordinate $\y$. The inversion symmetry is somewhat reminiscent of the string $S$-duality, which is known to be a self-duality of the string UV completion of  LD spacetime. We show that the dilatation symmetry uniquely defines the LD geometry, the exact requirement is that the line element be homothetic to itself under dilatation.

We study quantum fields living on the LD background. Following the principles of effective field theory we  construct an interactive, massive QFT on the LD background that respects  the dilatation and inversion symmetry. We find that the mass term for a scalar field has a negative lower bound. This bound is similar  to the Breitenlohner-Freedman bound from AdS, but the proof is closer in spirit to a flat space argument. 

We find that the scalar propagator in LD is related to a $(d+1)$-dimensional flat space propagator for a scalar with specific mass, up to $r$-dependent scaling factors. We analogously find that all the correlators of our EFT have a very peculiar structure, they are proportional to $(d+1)$-dimensional flat space correlators with $r$-dependent scaling factors in the external legs.

In order to study the holography of the LD spacetime, we place a flat brane parallel to the singularity and compute the boundary quantum effective action on it. The brane splits the spacetime into two inequivalent regions, LD$_\pm$. 

We study the 2-pt functions for a matter scalar and graviton in LD$_\pm$.  The correlators in each regions are almost identical, except that the spectrum in the LD$_-$ region features an isolated mode. { The similarity of the LD$_+$ and LD$_-$ correlators is a manifestation of the inversion symmetry of the background. The existence of the isolated mode in LD$_-$   shows that the singularity repels quantum fluctuations, {i.e.}~it acts as a (very smooth) boundary for the fields. }
{The  LD$_+$ region features no isolated  mode, which implies that gravity decouples at low-energy. This reflects the gravity decoupling expected in LST.}
We find that the 2-pt correlators always feature a gapped continuum for any $d$. This implies that the putative dual theory living on the brane, defined through the boundary quantum effective action, has a gapped spectrum. 

We evaluate simple scalar 3-pt and 4-pt holographic correlators built from the boundary quantum effective action. These correlators share peculiar features that also generalize to more complex diagrams: They develop poles at momentum space configurations corresponding to simultaneous kinematic thresholds. Furthermore the associated residue is proportional to a $(d+1)$-dimensional flat space $S$-matrix amplitude. Focusing on the main singularity, we  show that both of these phenomena occur for any diagram, and make it clear that these are consequences of the dilatation symmetry of the EFT.

We point out that the peculiar properties of the LD holographic correlators are reminiscent of those of a flat space toy-model {used to understand the singularity structure of 
the coefficients of the wavefunction of the Universe.}  The analogy implies that  bootstrap techniques developed for the wavefunction of the Universe can be transposed to the holographic correlators of the LD spacetime. An important difference is that singularities of the wavefunction coefficients are unphysical while those of the LD holographic correlators are physical. Since such singularities essentially define the correlators, they may even be used to understand/define the putative $d$-dimensional dual theory,  if it exists.
Overall, the perturbative correlators seem to tell us that the holographic theory is gapped. This idea matches the result for single-trace $T\overline T$-deformed CFT$_2$  derived via string techniques in  Refs.\,\cite{Asrat:2017tzd,Giveon:2017nie,Giribet:2017imm}.

We then take a completely different viewpoint on LD holography by putting the spacetime at finite temperature. Namely, we solve the $(d+1)$-dimensional field equations in the presence of a planar black hole ({i.e.}~black brane). We then project the bulk physics onto the brane, which generates the $d$-dimensional effective Einstein equation with a nontrivial stress tensor. The effective stress tensor is the  manifestation  of the bulk horizon on the brane. It receives contributions from both the dilaton bulk stress tensor and from the Weyl curvature projected onto the brane. The pressure from both terms cancel, such that the effective stress tensor matches the one of a {\it pressureless} perfect fluid at rest. Summarized: 
\be \boxed{ {\rm \it\,~Linear~Dilaton~Black~Hole}~(d+1\,  {\rm \it dim.} ) \quad \Leftrightarrow \quad  {\rm \it Pressureless~Fluid}~(d\,  {\rm \it dim.} ) } \nn \,\ee 

We test this remarkable result by letting the brane evolve with time in the bulk, hence defining  a $(d+1)$-dimensional linear dilaton braneworld. The evolution of the stress tensor is consistent with pressureless matter. We further check that the $(d+1)$-dimensional conservation equation projected onto the brane is satisfied.
The emergence of a pressureless fluid is reminiscent of a gapped spectrum --- since it is the expected behavior for a fluid of massive, nonrelativistic matter. We thus obtain another hint  that the dual $d$-dimensional theory, if it exists, is gapped. This hint, being at the level of the classical gravity solution, is completely independent of our study of the holographic correlators.

We study the thermodynamics of  the holographic fluid.  
{We find that $ E_{\rm fluid}\propto S_{\rm fluid}$ and that the proportionality constant is a universal (Hagedorn) temperature. }
These results match  the Hagedorn behavior found in LST and in  CFT$_2$ at large $T\overline T$ deformation.  The Hagedorn behavior appears for any $d$ in our low-energy approach. 
{Moreover, we find that the entropy $S_{\rm fluid}$ obtained from brane thermodynamics exactly reproduces the Bekenstein-Hawking entropy of the planar black hole.}
Summarized:
\be
\boxed{LD \ BH \  Thermodynamics\ (d + 1 \,  dim.) \quad \Leftrightarrow \quad Hagedorn \ Thermodynamics\ (d  \, dim.)\nonumber}
\ee

We close by pointing out a few open questions and directions. 
 
The computation of the holographic fluid features beautiful cancellations that lead to the vanishing  pressure. 
{This vanishing pressure is, to some extent, tied to the conformal dilatation symmetry of the background. It would be good to explore in details the interplay between the dilatation symmetry and the properties of the holographic fluid. }

The LD holographic correlators deserve  more study. In particular it would be good to analyze them via the bootstrap techniques developed for the wavefunction of the Universe. It would also be interesting to see if the LD correlators admit the polytope representation developed in \cite{Arkani-Hamed:2017fdk,Arkani-Hamed:2018bjr,Benincasa:2019vqr}, analogously to the flat space wavefunction toy-model.

{Here we have studied  holography on a brane parallel to the singularity, {i.e.}~on a timelike surface. Even though we obtain compelling results, another interesting direction could be to study holography on  the conformal null boundary of LD, similarly to flat space holography. 
This is a distinct analysis that we leave for future work. }

Finally the deep, overarching question is whether  there is an independent formulation of the $d$-dimensional dual theory that reproduces the holographic results obtained in this work. Is there,  in analogy with AdS$_{d+1}$/CFT$_d$,   an explicit dual to LD$_{d+1}$? 
While the existence of LST for $d = 6$ and the CFT$_2$ models may be encouraging signals,  the mysteries  of holography beyond AdS  mostly remain to be % uncovered. 
pierced.

\begin{acknowledgments}

We would like to thank S. Barbosa, G. von Gersdorff and L.L. Salcedo for useful discussions and comments. The work of EM is supported by the project PID2020-114767GB-I00 and by the Ram\'on y Cajal Program under Grant RYC-2016-20678 funded by MCIN/AEI/10.13039/ 501100011033 and by ``FSE Investing in your future'', by the FEDER/Junta de Andaluc\'{\i}a-Consejer\'{\i}a de Econom\'{\i}a y Conocimiento 2014-2020 Operational Programme under Grant A-FQM-178-UGR18, by Junta de Andaluc\'{\i}a under Grant FQM-225, and  by the ``Pr\'orrogas de Contratos Ram\'on y Cajal'' Program of the University of Granada. The work of MQ is partly supported by Spanish MICIN under Grant PID2020-115845GB-I00, and by the Catalan Government under Grant 2021SGR00649. IFAE is partially funded by the CERCA program of the Generalitat de Catalunya.
\end{acknowledgments}

\appendix

\section{General Solutions of the Dilaton-Gravity System}
\label{app:solutions}

We present in this appendix the most general solutions of the field equations at zero temperature~\eqref{eq:EoM1}-\eqref{eq:EoM2}, and in the presence of a planar black hole~(\ref{eq:EoMT1})-(\ref{eq:EoMT3}).

\subsection{Solution with No Black Hole}
\label{sec:app_T0}

We here consider the solution of the field equations for a metric in the absence of a black hole.
\subsubsection{Field equations}

The warped metric at zero temperature is
\be
ds^2 = e^{-2A(\y)}\eta_{\mu\nu} dx^\mu dx^\nu+ e^{-2B(\y)}d\y^2\,.  \label{eq:ds2_T0_App}
\ee
The $B(r)$ function can be freely chosen due to the freedom of redefinition in $\y. $  For instance, $B(r)$ can be absorbed by the change of variable 
$e^{-B(r)} dr = d\tilde r $,
which corresponds to proper coordinates. Alternatively, the change of variable $e^{-B(r)} dr = e^{-A(\tilde r)} d\tilde r$ leads to conformal coordinates. 
Here, however,  we need to  keep  arbitrary $B(r)$ to find consistent solutions when solving in the presence of a brane.

The field equations  are given by \eqref{eq:EoM1}-\eqref{eq:EoM2} and a third  equation. The three field equations are 
\begin{eqnarray}
\hspace{-0.3cm}&& A^{\prime\prime}(r) + A^\prime(r) B^\prime(r) - \bar \phi^\prime(r)^2 = 0 \,,  \label{eq:aEoM1}  \\
\hspace{-0.3cm}&&A^\prime(r)^2 + \frac{1}{D-1} \left(   2 e^{-2 B(r)} \bar 
V(\bar\phi) - \bar\phi^\prime(r)^2 \right) = 0 \,,    \label{eq:aEoM2}
\\
\hspace{-0.3cm}&& \bar\phi^{\prime\prime}(r) + \left( B^\prime(r)  - (D-1) A^\prime(r)  \right) \bar\phi^\prime(r) -  e^{-2B(r)} \frac{\partial \bar V}{\partial \bar\phi} = 0 \,. \label{eq:aEoM3}
\end{eqnarray}
Equation~(\ref{eq:aEoM3}) can be eliminated in favor of Eqs.~(\ref{eq:aEoM1})-(\ref{eq:aEoM2}) by means of the identity
\begin{eqnarray}
\frac{1}{(D-1)} \bar\phi^\prime(r) \cdot [\textrm{\ref{eq:aEoM3}}] &=&  A^\prime(r)  \cdot [\textrm{\ref{eq:aEoM1}}] - \left( B^\prime(r) + \frac{1}{2} \frac{d}{dr} \right)  [\textrm{\ref{eq:aEoM2}}] \,, \label{eq:identity}
\end{eqnarray}
so that there are two independent differential equations left. The number of integration constants in Eqs.~(\ref{eq:aEoM1})-(\ref{eq:aEoM3}) seem to be {four} by naive counting. {However Eq.~(\ref{eq:identity}) provides an algebraic relation between all the integration constants, which constrains their number to three.}

An alternative analysis of the system  (\ref{eq:aEoM1})-(\ref{eq:aEoM3}) can be done as follows. {For a given value of the bulk potential $\bar V(\bar\phi)$,} those equations can be written in terms of a superpotential $W(\phi)$  as
\begin{equation}
A^\prime(r) = \frac{1}{D-3} e^{-B(r)} \bar W(\bar\phi(r))  \,, \qquad \bar\phi^\prime(r) = \frac{1}{D-3} e^{-B(r)} \bar W^\prime(\bar\phi(r))  \,,   \label{eq:A_phi_W}
\end{equation}
and
\begin{equation}
\bar V(\bar\phi) = \frac{1}{2(D-3)^2} \left( \bar W^\prime(\bar\phi)^2 - (D-1) \bar W^2(\bar\phi) \right) \,, \label{eq:VW}
\end{equation}
where $\bar W \equiv W / [(D-2) M_D^{D-2}]$ {is the solution of Eq.~(\ref{eq:VW})}. These constitute three equations of first order, thus making it clear that there are \textit{three} integration constants, {a feature that has already been discussed in the literature~\cite{DeWolfe:1999cp}}.

\subsubsection{Integration constants and singularity}
\label{subsec:int_cte_sing}

Given the potential $\bar V(\bar\phi)$ of Eq.~(\ref{eq:V}), the solution of Eq.~(\ref{eq:VW}) is expressed as
\begin{equation}
\bar W(\bar\phi) = -(D-3) k \, e^{\bar\phi} \left[ 1 + \mathcal F\left( e^{(D-2) \bar\phi} \right) \right] \,,
\end{equation}
where $\mathcal F(x)$ is the solution of a first order differential equation. It thus contains an integration constant that we will denote by $s$. By using the method explained in Refs.~\cite{Megias:2018sxv,Papadimitriou:2007sj}, this equation can be solved by making an expansion in $s$. This produces the following leading term
\begin{equation}
\mathcal F(x) = x  s + O(s^2) \,.
\end{equation}

In the following we  assume that {$\bar V(\bar\phi) \stackrel{\bar\phi \to \infty}{\sim} \bar W^2(\bar \phi)$}  which corresponds to a good  singularity (see {e.g.}~Refs.~\cite{Cabrer:2009we,Megias:2014iwa} for a discussion), and set the integration constant to $s = 0$. {This removes one integration constant from the solutions, leaving only two.}

\subsubsection{Solutions}
\label{subsec:sol_T0}

A convenient way to derive the general solution of the system (\ref{eq:aEoM1})-(\ref{eq:aEoM3}) is to assume that $B(r)$ is linearly related to $A(r)$, {i.e.}~$B(r) = \alpha_1 A(r) + \alpha_2$.  Then, after setting $s=0$ the system admits the following solution, 
\begin{eqnarray}
&&A(r) = - a \log\left( \frac{\y - \y_0}{L} \right)  \,, \quad B(r) = -(a-1) \log\left( \frac{\y - \y_0}{L} \right) - \log \C 
\,, \label{eq:Ar_Br} \\
&&\bar \phi(r) = - a \log \left(  \frac{\y - \y_0}{L} \right) + \log\left( \frac{ |a|}{k L \C } \right)   \,, \label{eq:phir}
\end{eqnarray}
{where we have defined the parameters $a$ and $c$ as $\alpha_1 \equiv (a-1)/a$ and $\alpha_2 \equiv -\log c$.  These solutions depend on a set of four parameters: {two integration constants $(L,r_0)$ and two constants $(a,c)$;} which is certainly reducible. The domains of these parameters are $a\in\mathbb{R}_{/\{0\}}$ and $L,\C > 0$. {Notice that the integration constant $L$ is equivalent to an additive constant $c_A$ in $A(r)$, as it can be seen by expressing it as $L = e^{c_A}/k$ in addition to $c = e^{a c_A} \bar c$.} 
In fact fixing $c_A= - \bar v_b$ leads to the fixing $1/L = \eta \equiv k \, e^{\bar v_b}$, which is usually assumed in warped models~\cite{Megias:2021mgj}.

Among the four parameters, $r_0$, $a$ and $\C$ are redundancies of the description of the $D$-dimensional geometry, because a change in the value of either of them is equivalent to a diffeormorphism.  First, a change in the value of $r_0$ corresponds to a shift  in the coordinate~$r$.  We thus set $r_0 = 0$ without loss of generality. Second, considering the resulting line element
\begin{align}
ds^2_a(x^\mu_a,r_a)&=\left(\frac{r_a}{L} \right)^{2a}\eta_{\mu \nu}dx^\mu_a dx^\nu_a + \C^2 \left(\frac{r_a}{L} \right)^{2(a-1)} dr_a^2 \,,
\label{eq:dsa2}
\end{align}
the change of coordinates from $(x_a^\mu,r_a)$ to $(x_b^\mu,r_b)$ 
\begin{equation}
    \frac{r_b}{L}=\left| \frac{b}{a} \right|^{\frac{1}{b}}\left(\frac{r_a}{L} \right)^{\frac{a}{b}} \,, \quad x_b^\mu= \left|\frac{a}{b}\right|x_a^\mu \,\label{eq:a_b}
\end{equation} 
with $a, b \neq 0$, gives
\be 
ds^2_a(x^\mu_a,r_a)=ds^2_b(x^\mu_b,r_b)\,\quad\quad \bar\phi_a(r_a) = \bar\phi_b(r_b) \,, \label{eq:phi_a_b}
\ee
where
\begin{equation}
    \bar\phi_a(r_a) \equiv -a\log\left(\frac{r_a}{L}\right) +\log\left(\frac{|a|}{kL \C}\right)  \,. \label{eq:phi_a}
\end{equation}
Therefore the transformation Eq.~\eqref{eq:a_b} is equivalent to a change in the parameter $a$. 
Third, given two solutions with parameters $\C$ and $\tilde \C$, one can similarly check that the map
\begin{equation}
\tilde r = \left( \frac{\C}{\tilde \C} \right)^{\frac{1}{a}} r \,, \qquad \tilde x^\mu = \frac{\tilde \C}{\C} x^\mu \, \label{eq:transf_CB}
\end{equation}
gives $ds^2 = d\tilde s^2$ and $\bar\phi(r) = \tilde{\bar{\phi}}(\tilde r)$.  Therefore the map Eq.\,\eqref{eq:transf_CB} is equivalent to a change in the parameter $\C$. 
In summary, $a$ and $\C$ parametrize redundancies in the decription of the geometry and can thus be fixed. 
It follows that only \textit{one} combination of parameters may have a physical meaning. 

In order to figure out which combination of parameters exactly is physical, it is instructive to compute  the mass gap of the continuum. We focus on the graviton. 
Using the conformal coordinates $z= \C L \log\left( \frac{L}{\y} \right)$, we  can   determine the mass gap from  the effective Schr\"odinger potential,  
 % s
 \begin{equation}
 V_{\sigma}(z) = \frac{(d-1)^2}{4} A^\prime(z)^2 - \frac{d-1}{2} A^{\prime\prime}(z) = \left( \frac{d-1}{2}  \sigma \right)^2  \equiv m_g^2 \,. 
 \end{equation}
 The $m_g$ parameter  is identified as the mass gap. 
 We have thus \be
m_g = \frac{d-1}{2} \sigma \qquad \textrm{with} \qquad \sigma \equiv \frac{a}{\C L}\,. \label{eq:sigma_scale}
 \ee
 We can see that the physical quantity $m_g$ is related to a combination of the three parameters.  
  In sections \ref{se:fields} and \ref{se:correlators} our ``gauge-fixing" is $a=1$, $\C=1$, in which case the $L$ parameter is physical, with $m_g = \frac{d-1}{2L}$.

\paragraph{Comparison to Ref.~\cite{Megias:2021mgj}.}
 At this point it would be useful to make connection with the notation used in Ref.~\cite{Megias:2021mgj} for the case $D=5$. Using  the metric ansatz of~\cite{Megias:2021mgj}
\begin{equation}
ds^2 = e^{-2A(z)} \left( \eta_{\mu\nu} dx^\mu dx^\nu + dz^2 \right) \,, \label{eq:ds2_z}
\end{equation}
with $z= \C L \log\left( \frac{L}{\y} \right)$,
 the solution is found to be
\begin{equation}
A(z) = \sigma z   \,, \qquad  \bar \phi(z) = \sigma z + \log\left( \frac{|\sigma|}{k} \right)  \, \label{eq:Aphi_z}
\end{equation}
with the scale $\sigma$ defined in Eq.~(\ref{eq:sigma_scale}).
Here $\sigma$ can have either sign, and the curvature singularity is in the limit $z \to \pm\infty$ for~$a \gtrless 0$.~\footnote{We could also have defined the conformal coordinate $z$ and scale $\sigma$ as 
\begin{equation}
z= \textrm{sgn}(a) \C L  \log\left( \frac{L}{\y} \right) \,, \qquad \sigma \equiv \frac{|a|}{\C L}\,. 
\end{equation}
With this definition, $\sigma$ is always positive and the singularity is in the limit $z \to +\infty$.}

\subsubsection{Fixing the constants}
\label{sec:fixing}

The remaining parameters are fixed as follows.
\begin{itemize}
\item[\it i)] {\it No brane}: We set $a=1$, $\C = 1$ with no loss of generality. The solution turns out to be
\begin{equation}
ds^2 = d\y^2+\frac{\y^2}{\L^2}\eta_{\mu\nu}dx^\mu dx^\nu \,,  \qquad \bar \phi(r) = -\log \left( k\y\right)  \,,
\end{equation}
and the scale $\sigma$ is $\sigma = 1/L$. Any other value of $a,\C$ can be recovered by a coordinate transformation.

\item[\it ii)] {\it Brane at $r=r_b$}: {Fixing $a=1$}, the fact that the value of the scalar field at the brane is independent on $r_b$ implies (see Sec.~\ref{se:conditions} for other equivalent conditions)
\begin{equation}
\C = \frac{1}{e^{\bar v_b} k r_b}  \,. \label{eq:eCB}
\end{equation}
The solution turns out to be
\begin{equation}
ds^2 =  \frac{1}{\eta^2 r_b^2} d\y^2 + \frac{\y^2}{\L^2}\eta_{\mu\nu}dx^\mu dx^\nu \,,  \qquad \bar \phi(r) = \log \left( \frac{\y_b}{\y} \right) + \bar v_b  \,,
\end{equation}
where the $r_b$-independent scale $\eta$ is given by
\begin{equation}
\eta \equiv k \, e^{\bar v_b} \,.  \label{eq:app_eta}
\end{equation}
The scale $\sigma$ turns out to be
\begin{equation}
\sigma = \frac{1}{\C L} = \eta \frac{r_b}{L} \,.
\label{eq:etabar_eta}
\end{equation}
\end{itemize}

\subsection{Planar Black Hole solution}
\label{sec:app_T}

Here we will solve the field equations of the metric in the presence of a blackening factor.

\subsubsection{Field equations}

The metric in the presence of a planar black hole is
\be
ds^2= e^{-2A(\y)} \left( -f(r) d\tau^2 + d\x_{D-2}^2 \right) + \frac{e^{-2B(\y)}}{f(\y)} d\y^2\,. \label{eq:ds_BH}
\ee
Due to the breaking of the $(D-1)$-Lorentz invariance by the black hole, there is one more field equation than in the zero temperature case. 

The field equations are given by \eqref{eq:EoMT1}-\eqref{eq:EoMT3} and a fourth equation. The four field equations are
\begin{eqnarray}
\hspace{-0.3cm}&& \frac{f^{\prime\prime}(r)}{f^\prime(r)} + B^\prime(r) - (D-1) A^\prime(r) = 0 \,,  \label{eq:aEoMT1}  \\
\hspace{-0.3cm}&& A^{\prime\prime}(r) + A^\prime(r) B^\prime(r) - \bar \phi^\prime(r)^2 = 0 \,,  \label{eq:aEoMT2}  \\
\hspace{-0.3cm}&&A^\prime(r)^2 + \frac{1}{D-1} \left(  2 \frac{e^{-2 B(r)}}{f(r)} \bar 
V(\bar\phi) - \frac{f^\prime(r)}{f(r)} A^\prime(r) - \bar\phi^\prime(r)^2  \right) = 0 \,,    \label{eq:aEoMT3}
\\
\hspace{-0.3cm}&& \bar\phi^{\prime\prime}(r) + \left( B^\prime(r) + \frac{f^\prime(r)}{f(r)}  - (D-1) A^\prime(r)  \right) \bar\phi^\prime(r) -  \frac{e^{-2B(r)}}{f(r)} \frac{\partial \bar V}{\partial \bar\phi} = 0 \,. \label{eq:aEoMT4}
\end{eqnarray}
 Eq.~(\ref{eq:aEoMT4}) can be eliminated in favor of Eqs.~(\ref{eq:aEoMT1})-(\ref{eq:aEoMT3}) by means of the identity
\begin{eqnarray}
2 f(r) \bar\phi^\prime(r) \cdot [\textrm{\ref{eq:aEoMT4}}] &=&  -A^\prime(r) f^\prime(r) \cdot [\textrm{\ref{eq:aEoMT1}}] + \left( 2(D-1) f(r) A^\prime(r) - f^\prime(r) \right) \cdot [\textrm{\ref{eq:aEoMT2}}] \nonumber \\
&&- (D-1) \left( f^\prime(r) + 2 f(r) B^\prime(r) + f(r)  \frac{d}{dr} \right) [\textrm{\ref{eq:aEoMT3}}] \,. \label{eq:aidentityT}
\end{eqnarray}
Then, there are three independent differential equations left,~\footnote{An analysis of these equations with $B(r) = A(r)$ has been done in {e.g.}~Ref.~\cite{Gursoy:2008za}.} {with {\it five} integration constants}

{As in the solution with no black hole,} this system of equations can also be written as an equivalent system in terms of a superpotential at finite temperature, $\bar W(\bar\phi)$~\cite{Megias:2010ku}. This is given by
\begin{align}
0 &= \frac{f^{\prime\prime}(r)}{f^\prime(r)} + B^\prime(r) - \frac{D-1}{D-3} e^{-B(r)} \bar W(\bar\phi(r))  \,,  \\
\bar V(\bar\phi) &= \frac{f}{2(D-3)^2} \left( \bar W^\prime(\bar\phi)^2 - (D-1) \bar W^2(\bar\phi) + (D-3) \frac{f^\prime}{f} e^{B} \bar W(\bar\phi) \right) \,,
\end{align}
in addition to the two equations in~(\ref{eq:A_phi_W}). The system is formed by three differential equations of first order, and one differential equation of second order, thus leading to \textit{five} integration constants.

\subsubsection{Solutions}

We follow a similar analysis to that in Secs.~\ref{subsec:int_cte_sing} and~\ref{subsec:sol_T0}. First, we demand a good  singularity and   set $s=0$ which removes one integration constant, leaving four. We find that the system admits the solution
\begin{eqnarray}
&&A(r) =  - a \log\left( \frac{\y - \y_0}{L} \right)  \,,  \label{eq:AT} \\ 
&&B(r) = -(a-1) \log\left( \frac{\y - \y_0}{L} \right) - \log \C + \log c_f 
\,,  \label{eq:BT}\\
&&f(r) = \frac{1}{c_f^2} \left[ 1 - \left( \frac{\y_h - \y_0}{\y - \y_0}\right)^{a(D-2)}  \right] \,, \label{eq:fT} \\
&&\bar \phi(r) = - a \log \left( \frac{\y - \y_0}{L} \right)  + \log\left( \frac{|a|}{k L \C } \right)  \,, \label{eq:phibT}
\end{eqnarray}
with the superpotential given by 
\begin{equation}
\bar W(\bar\phi) = - c_f (D-3) k \, e^{\bar\phi} \,.
\end{equation}

The solution has a set of six parameters: {four integration constants $(L, r_0, r_h,c_f)$ and two constants $(a,c)$.}  As in the zero temperature case,  $r_0$, $a$, and $c$ are redundancies in the description of the geometry that can be gauge-fixed.  We readily set $r_0=0$. The change in~$a$ corresponds to 
the coordinate transformation  of Eq.~(\ref{eq:a_b}) together with the redefinition
\begin{equation}
    \frac{r_{b,h}}{L}=\left| \frac{b}{a} \right|^{\frac{1}{b}}\left(\frac{r_{a,h}}{L} \right)^{\frac{a}{b}} \,,
\end{equation}
where $r_{a,h}$ is the horizon location in the $r_a$ coordinate.  The same conclusion is obtained when considering the transformation of Eq.~(\ref{eq:transf_CB}) together with the redefinition~$\tilde r_h = \left( \C/\tilde \C\right)^{\frac{1}{a}}  r_h$. {Finally, the constant $c_f$ is another redundancy in the geometry, as a change in $c_f$ corresponds to the coordinate transformation
\begin{equation}
\tilde t = \frac{\tilde c_f}{c_f} t \,,
\end{equation}
leading to $ds^2 = d\tilde s^2$, and can thus be fixed. We will consider the value $c_f = 1$ to guarantee that the black hole solution tends to the metric of Eq.~(\ref{eq:ds2_T0_App}) in the $r_h\to 0$ limit.}

\subsubsection{Temperature of a planar black hole}

\label{se:Th}

{The temperature of a planar horizon can be obtained by the standard method of demanding the absence of  conical singularity. Given the Euclidean time $(\tau_E \equiv i \tau)$ version of the black hole metric of Eq.~(\ref{eq:ds_BH}), we study the behavior of $ds^2$ near the horizon, where $f(r_h) = 0$. The series expansion of the variables at the first non-vanishing order is
  \begin{align}
    A(r) & = A(r_h) + \mathcal O\left(r-r_h\right) \,, \\
    B(r) & = B(r_h) + \mathcal O\left(r - r_h \right) \,, \\
    f(r) & = f^\prime(r_h) (r - r_h) +\mathcal O\left((r-r_h)^2\right)\,.
  \end{align}
  At this point it is useful to define the new variable $\xi$ as  $r = r_h (1 + \xi^2)$, in terms of which the line element near the horizon writes
  \begin{equation}
  ds^2 \simeq e^{-2A(r_h)} \left( r_h f^\prime(r_h) \xi^2 d\tau_E^2 + e^{2(A(r_h) - B(r_b))} \frac{4 r_h}{f^\prime(r_h)} d\xi^2 + d\x_{D-2}^2  \right) \propto d\xi^2 + \xi^2 d\theta^2 + \cdots \,,
  \end{equation}
where we have defined $\theta \equiv \frac{1}{2} e^{B(r_h) - A(r_h)} |f^\prime(r_h)| \tau_E$. This is the usual expression of the line element in polar coordinates. The conical singularity is absent if the line element has periodicity $\theta \to \theta + 2\pi$. This translates into a periodicity in Euclidean time of the form $\tau_E \to \tau_E + \beta$ with $\beta \equiv 1/T_h$. We finally arrive at the expression of the temperature
\begin{equation}
T_h = \frac{1}{4\pi} e^{B(r_h) - A(r_h)} |f^\prime(r_h)|  \,. \label{eq:Th_general}
\end{equation}
This expression is compatible with previous results reported in the literature (see {e.g.}~Ref.~\cite{Megias:2018sxv} for a metric with $B(r) = 0$). 
}

\subsubsection{Temperature and entropy of the LD  black hole}

Eq.~(\ref{eq:Th_general}) leads to the following result in the LD model
\begin{equation}
T_h = \frac{D-2}{4\pi} |\sigma|  \qquad \textrm{with} \qquad \sigma = \frac{a}{\C L}   \,.\label{eq:aTh}
\end{equation}
On the other hand, the entropy  of the black hole can be computed by using the Bekenstein-Hawking entropy formula
\begin{equation}
S_h = \frac{\mathcal A}{2 G_D}  \,,
\end{equation}
where $G_D \equiv \frac{M_D^{2-D}}{8\pi}$ is the $D$-dimensional Newton constant, and we have introduced an extra factor of $2$ from the $\mathbb{Z}_2$ orbifold convention adopted in Sec.~\ref{se:finiteT}.  By using the metric of Eq.~(\ref{eq:ds2BH}), the  area of the event horizon is computed as
\begin{equation}
\mathcal A = \int d^{D-2} x \sqrt{ |\bar g| }  = \int d^{D-2}x  \, e^{-(D-2) A(r_h) }  =  V_{D-2} \left(\frac{r_h}{L} \right)^{a(D-2)}\,.
\end{equation}
The entropy density is then given by
\begin{equation}
s_h \equiv \frac{S_h}{V_{D-2}} = \frac{1}{2 G_D} \left(  \frac{r_h}{L} \right)^{a (D-2)} \,.
\end{equation}
The zero temperature solution is recovered by taking the limit $r_h^{a} \to 0$, which means  moving the horizon towards the singularity.

\subsubsection{Fixing the constants}
\label{subsec:fixing}

We fix the remaining parameter similarly to the case with no black hole.

\begin{itemize}
 \item[\it i)] {\it No brane}: We set $a=1$, $\C = 1$ with no loss of generality. The solution turns out to be
\begin{equation}
ds^2 = \frac{d\y^2}{f(r)} + \frac{\y^2}{\L^2}\left( -f(r) d\tau^2 + d\x_{D-2}^2 \right) \,,
\qquad \bar \phi(r) = -\log \left( k\y\right) \,,
\end{equation}
with
\begin{equation}  
f(r) = 1 - \left( \frac{r_h}{r} \right)^{D-2} \,.
\end{equation}
The horizon temperature is given by
\begin{equation}
T_h = \frac{D-2}{4\pi L} \,.
\end{equation}

\item[\it ii)] {\it Brane at $r=r_b$}: Applying the same conditions as in the case of no black hole, the solution is
\begin{equation}
ds^2 =  \frac{1}{\eta^2 r_b^2} \frac{d\y^2}{f(r)} + \frac{\y^2}{\L^2} \left( -f(r) d\tau^2 + d\x_{D-2}^2 \right) \,, \qquad  \bar \phi(r) = \log \left( \frac{\y_b}{\y} \right) + \bar v_b  \,,
\end{equation}
with
\begin{equation}
f(r) = 1 - \left( \frac{r_h}{r}\right)^{D-2} 
\end{equation}
where $\eta$ is given by Eq.~(\ref{eq:app_eta}). %As it happened in the zero temperature solution, the black hole solution with $\C = 0$ would still be valid, although in this case $\bar v_b$ would depend on $r_b$ according to Eq.~(\ref{eq:eCB}).
The temperature is then
\begin{equation}
T_h = \frac{D-2}{4\pi} \sigma \,,
\end{equation}
where $\sigma$ is given by Eq.~(\ref{eq:etabar_eta}).

\end{itemize}

\bibliographystyle{JHEP}
\bibliography{biblio}

\end{document}